\documentclass[acmsmall,screen]{acmart}
\AtBeginDocument{%
  }
    
\usepackage[ruled,linesnumbered]{algorithm2e}
\usepackage{multirow,multicol}
\setcopyright{acmcopyright}
\copyrightyear{2023}
\acmYear{2023}
\acmDOI{XXXXXXX.XXXXXXX}
\newcommand\Tstrut{\rule{0pt}{2.6ex}}

\usepackage{graphicx}
\acmPrice{}
\acmISBN{}

\begin{document}

\title{Generating Robust Adversarial Examples against Online Social Networks (OSNs)}


\author{Jun Liu}
\email{yc07453@umac.mo}
\affiliation{%
  \institution{University of Macau}
  \city{Macau}
  \country{China}
}
\author{Jiantao Zhou}
\authornote{Corresponding author}
\email{jtzhou@umac.mo}
\affiliation{%
  \institution{University of Macau}
  \city{Macau}
  \country{China}
}

\author{Haiwei Wu}
\affiliation{%
  \institution{University of Macau}
  \city{Macau}
  \country{China}}
\email{yc07912@umac.mo}

\author{Weiwei Sun}
\affiliation{%
  \institution{Alibaba Group}
  \city{Hangzhou}
  \country{China}}
\email{sunweiwei.sww@alibaba-inc.com}

\author{Jinyu Tian}
\affiliation{%
  \institution{Macau University of Science and Technology}
  \city{Macau}
  \country{China}}
\email{jytian@must.edu.mo}

\renewcommand{\shortauthors}{Jun L et al.}

\begin{abstract}
 
Online Social Networks (OSNs) have blossomed into prevailing transmission channels for images in the modern era. Adversarial examples (AEs) deliberately designed to mislead deep neural networks (DNNs) are found to be fragile against the inevitable lossy operations conducted by OSNs. As a result, the AEs would lose their attack capabilities after being transmitted over OSNs. In this work, we aim to design a new framework for generating robust AEs that can survive the OSN transmission; namely, the AEs before and after the OSN transmission both possess strong attack capabilities. To this end, we first propose a differentiable network termed SImulated OSN (SIO) to simulate the various operations conducted by an OSN. Specifically, the SIO network consists of two modules: 1) a differentiable JPEG layer for approximating the ubiquitous JPEG compression and 2) an encoder-decoder subnetwork for mimicking the remaining operations. Based upon the SIO network, we then formulate an optimization framework to generate robust AEs by enforcing model outputs with and without passing through the SIO to be \emph{both} misled. Extensive experiments conducted over Facebook, WeChat and QQ demonstrate that our attack methods produce more robust AEs than existing approaches, especially under small distortion constraints; the performance gain in terms of Attack Success Rate (ASR) could be more than 60\%. Furthermore, we build a public dataset containing more than 10,000 pairs of AEs processed by Facebook, WeChat or QQ, facilitating future research in the robust AEs generation. The dataset and code are available at \url{https://github.com/csjunjun/RobustOSNAttack.git}.  
\end{abstract}

\ccsdesc[500]{Security and privacy~Social network security and privacy}
\ccsdesc[300]{Security and privacy~Privacy protections}
\keywords{Adversarial examples, adversarial images, robustness, online social networks, deep neural networks}


\maketitle

\section{Introduction}
{I}n recent years, Deep Neural Networks  (DNNs) have attracted enormous attentions and contributed to computer vision applications including image classifications\cite{krizhevsky2012imagenet,endo2020cnn}, object detections\cite{chen2020high}, inpainting \cite{pathak2016context}, etc. However, concurrent works revealed the vulnerability of DNNs by crafting adversarial examples (AEs)\footnote{Adversarial examples and adversarial images are interchangeably used in this work.}, which are close to clean images in human perception but could deceive DNNs with malicious perturbations \cite{qin2022gradually,goodfellow2014explaining,sun2022minimum,madry2017towards,dong2018boosting,carlini2017towards,chen2017zoo,brendel2018decision,
yatsura2021meta}.
Due to the exquisite distribution and small magnitude of the adversarial perturbations in nature, AEs are fragile to lose their attack capabilities when being transmitted/delivered over various lossy channels.

\begin{figure}[t!]
\centering
\includegraphics[width=0.5\textwidth]{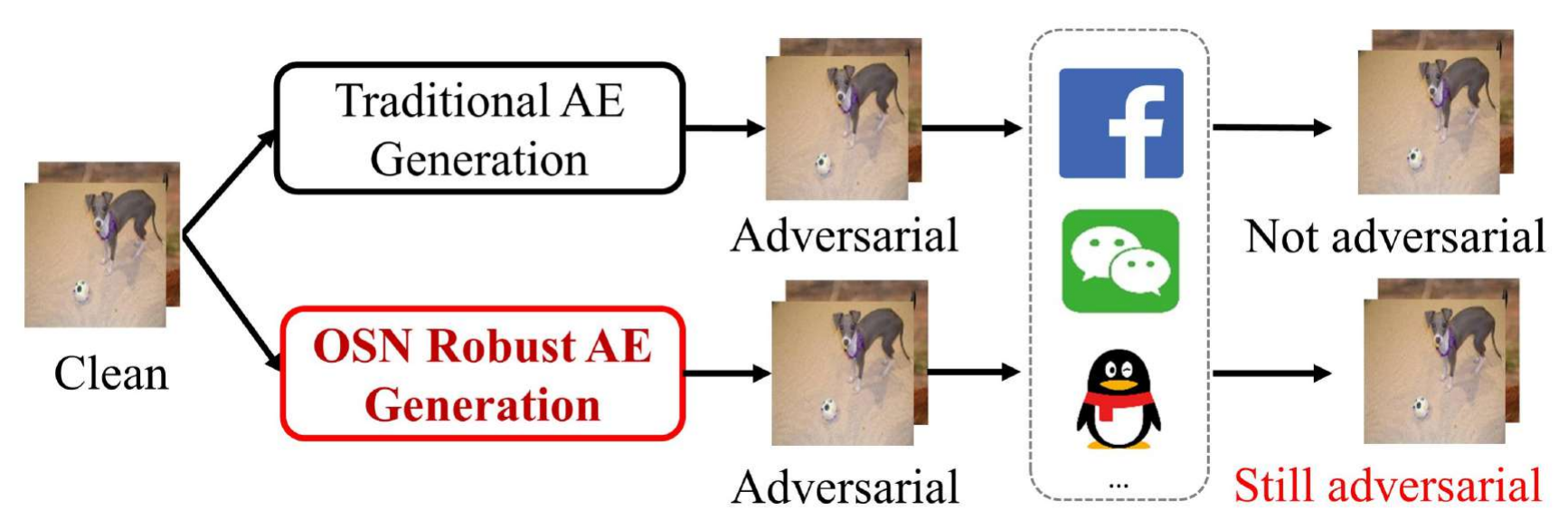}
\caption{The scenario of transmitting AEs over OSNs and the impact on their attack capabilities.}
\label{figstory}
\end{figure}

Many existing works \cite{athalye2018synthesizing,luo2018towards,wang2021exploring,eykholt2018robust,tsai2020robust} have attempted to produce robust AEs against a variety of degradations in the physical world such as viewpoints, printers, rotations, camera noise, lighting conditions, etc., or lossy operations in the digital domain, e.g., compression \cite{9893158,shi2021generating,shin2017jpeg,wang2020towards,zhang2023low}. Nevertheless, little research has considered the influence of Online Social Networks (OSNs) \cite{hu2016budget} on the effectiveness of AEs. In fact, OSNs have blossomed into prevailing transmission channels for images in the modern era, and many user-generated contents on OSNs can be utilized to predict volunteerism, user interest, and etc \cite{10.1145/2832907,nie2022learning}. It has also been shown that the lossy image processing procedures inevitably performed on OSNs \cite{wei2016osn,zhang2019multiple,wei2020osn,sunoptimal,9686650,zhu2022image} do greatly damage the transmitted images, and could result in severe degradation of attack performance of AEs. Such a scenario is illustrated in Fig.\ref{figstory}.  


In this work, we propose a novel attack framework to generate robust AEs that can survive the lossy transmissions over OSNs. In other words, the AEs before and after the OSN transmission both are expected to possess strong attack capabilities. To this end, we need to deal with two critical problems: 1) figuring out the image processing procedures of OSNs and 2) designing a proper robust AEs generation method accordingly. The former problem is analogous to the channel modeling/estimation when designing an error-resilient communication system. Clearly, having a more precise knowledge on the OSN degradation models could significantly benefit the solution to the latter problem. There are some existing works \cite{sunoptimal,wei2020osn} revealing partial operations in OSNs, including resizing and JPEG compression. However, many operations, e.g., the enhancement filtering and postprocessing, are still unknown \cite{wei2020osn}. More importantly, it seems to be nontrivial to incorporate the knowledge of the OSN channel into the framework of generating robust AEs.  

To solve the first critical problem, we design a SImulated OSN (SIO) network to imitate the behaviors of OSNs. Our design goal is to force SIO to predict the OSN outputs as accurately as possible, and at the same time, make it easy to be concatenated with a target DNN model. More specifically, the proposed SIO consists of 1) an encoder-decoder based on a U-Net\cite{ronneberger2015u} with ``Squeeze-and-Excitation'' blocks \cite{hu2018squeeze,roy2018concurrent,roy2018recalibrating}, to enhance the learning ability for the unknown and intangible manipulations and 2) a differentiable JPEG layer for approximating the JPEG compression with various Quality Factor ($QF$) values. In addition, we devise a shortcut to the output of the U-Net for the residual learning, such that the network can better learn the OSN processing traces, rather than the image content itself. Such a differentiable, end-to-end network makes the gradient backpropagation with respect to images possible, and hence, it can be readily incorporated with any target DNN model.

For solving the second critical problem, we append the pre-trained SIO with the target DNN model. Our goal is to generate robust AEs which can maintain their attack capabilities upon the OSN transmission. In other words, whether the AEs are transmitted over OSNs or not, they are expected to stay effective in misleading the target DNN model. Accordingly, we design two constraints and incorporate them into the optimization framework of generating the AEs. We derive two types of solution for integrating the gradient of original inputs with that of OSN processed inputs; namely, \emph{the gradient projection solution} maximizing our proposed joint cross-entropy (CE) loss, and the \emph{Lagrange-form solution} minimizing a joint margin loss \cite{carlini2017towards}. It should also be noted that the proposed framework could accommodate any superior optimization approaches and different OSN simulation networks.

Extensive experimental results demonstrate that our proposed SIO network and attack framework significantly outperform vanilla attack methods and the state-of-the-art robust AE generation schemes \cite{shin2017jpeg,wang2020towards}, especially under small adversarial distortion constraints. Specifically, we choose Facebook as the major OSN platform for presenting the experimental results, since Facebook is one of the most representative and complicated OSNs with a large number of users \cite{niu2014multi,liu2015deep,tang2017traffic,trzcinski2017predicting,wu2010improving,zhao2016predicting,luo2017social,you2016sentiment}. It is also shown that our proposed scheme can be readily extended to other OSN platforms, e.g., WeChat and QQ, and achieves considerable performance gains. To facilitate future research on generating robust AEs over OSNs, we also build a public dataset containing more than 10,000 pairs of AEs transmitted over Facebook, WeChat or QQ.

Our contributions can be summarized as followings:
\begin{itemize}
\item We design a network, SIO, to simulate the image processing procedures of OSNs and achieve pretty good performance in mimicking OSNs, such as Facebook, WeChat and QQ.
\item Based upon SIO, we propose a new attack framework for generating robust AEs against transmission over given OSNs, which consists of novel loss functions and constraints. The generated AEs possess strong attack capabilities, before and after the OSN transmission. Our attack framework can also readily accommodate multiple optimization algorithms and OSN models.  
\item Extensive experiments demonstrate that our proposed attack framework produces more robust AEs over Facebook, WeChat and QQ than state-of-the-art competitors, especially in the case of small adversarial distortion constraints.
\item We build a public dataset of AEs transmitted over Facebook, WeChat or QQ for further research.
\end{itemize}

The rest of this paper is organized as follows. Section \ref{section:rewo} reviews related works on existing attack methods for generating AEs and robust ones, together with the manipulations of OSNs. Section \ref{section:simosn} introduces the simulation network design for OSNs. The details of our proposed attack framework are then given in Section \ref{section:genae}. Section \ref{section:expe} shows the performance of our proposed method and comparison with the competing algorithms in terms of attack success rate (ASR) over Facebook, WeChat or QQ. Finally, Section \ref{section:con} concludes.

\section{Related Works}\label{section:rewo}
\subsection{Adversarial Attack Methods}

Traditional adversarial attack methods for generating AEs can be roughly categorized into targeted and untargeted, depending on whether AEs' labels are pre-assigned or not. From the perspective of the attack model, adversarial attack methods can also be divided into black-box and white-box methods. In the black-box setting, the adversary can only get the output of the attack model, without knowing the internal status, e.g., the gradient information and the network weights. In the white-box setting, both the network architecture and weights of the attack model are fully accessible to the adversary. In this work, our focus is on the white-box attack, which can be roughly classified into two mainstreams including gradient projection \cite{goodfellow2014explaining,madry2017towards,dong2018boosting,rony2019decoupling} and Lagrange-form \cite{szegedy2014intriguing,carlini2017towards}, together with a few other attack methods \cite{moosavi2016deepfool,zhang2020walking}. Let us now briefly review these two mainstreams and other attacks below.

1) \textit{Gradient projection attacks} attempt to find a feasible AE with respect to the misclassification loss function under a distortion budget. This is achieved by utilizing gradient descent to search a local minimum and projection operations to control the perturbation distance. Along this line of research, Fast Gradient Sign Method (FGSM)\cite{goodfellow2014explaining} operated a one-step gradient descent to walk towards the direction of minimizing the negative CE loss. As improvements to FGSM, iterative methods perform gradient descent repeatedly. Projected gradient descent (PGD) \cite{madry2017towards} projected the perturbed image into an allowed range in each iteration. Then the Momentum Iterative Fast Gradient Sign Method (MIFGSM) \cite{dong2018boosting} extended PGD with the momentum method to accumulate a velocity vector in the gradient direction during iterations. In view of the constant step size towards the gradient in each iteration of the above iterative methods, Decoupled Direction and Norm (DDN) \cite{rony2019decoupling} adjusted the step size in each iteration according to whether the last perturbed image is adversarial or not, which resulted in smaller distortions than aforementioned approaches.

2) \textit{Lagrange-form attacks} consider the ASR and the distortion simultaneously by introducing both of their constraints into the optimization objective function, potentially achieving higher ASR and smaller adversarial perturbations. Moosavi-Dezfooli \emph{et al.} \cite{szegedy2014intriguing} formulated a Lagrange-form of the optimization problem for minimizing a combination of a success constraint and a distortion distance, and solved it by limited BFGS (L-BFGS) under a box constraint. Another well-known method C\&W \cite{carlini2017towards} eliminated the box constraint of the distortion by a change of variable and refined the success constraint so as to adjust a success margin by a parameter. It should be noted that the Lagrange-form attacks can hardly bind the incurred distortions, as the distortion term is incorporated into the optimization objective function.

There are also some \textit{other attacks} such as DeepFool \cite{moosavi2016deepfool} and Walking on the Edge (WE) \cite{zhang2020walking}. DeepFool directly minimizes the distance metric by approaching the minimal distance in the direction of normal vectors, until crossing the decision boundary that is linear in their hypothesis. WE \cite{zhang2020walking} optimized the perturbation magnitude by iteratively projecting the distortion gradient on the tangent space of the manifold of the decision boundary.

\subsection{Robust Adversarial Examples}
Above traditional attack methods focus on the AE generation in the ideal conditions, namely, no disturbances are imposed to the generated AEs. There are also a few recent works paying more attention to the robustness of AEs against transformations in the digital domain or physical world. Clearly, this is very important for the practical deployment of AEs. To fight against the distortion by JPEG compression, \cite{shin2017jpeg} generated JPEG-resistant AEs by including a differentiable and simple polynomial that approximates JPEG to the target model. Following a similar strategy, \cite{wang2020towards} used a DNN to approximate compression methods including JPEG for crafting AEs resisting unknown compression methods. Noticing that the largest errors in JPEG are caused by the rounding operation, \cite{shi2021generating} used the gradient of AEs in the DCT domain to guide the rounding so as to generate JPEG-robust AEs. Some other works utilized the intrinsic property of the target model to craft robust AEs. For example, \cite{luo2018towards} crafted generic perturbations by maximizing the gap between the logits of the target class and the ground truth. To the best of our knowledge, \cite{wang2020towards} is the only one that explicitly addressed the problem of generating robust AEs against OSNs. However, they modeled OSN as a pure compressor and used an approximation model named ComModel to imitate its behavior. As mentioned in many existing works \cite{wei2016osn, sunoptimal}, compression is only one part of OSN operations and there are many others having a great impact on the transmitted images as well. In fact, even for the same type of compression method (e.g., JPEG), different parameters (e.g., $QFs$) could lead to vastly different compressed images. Furthermore, if the generated AEs are actually not transmitted over OSNs, then sharp drops in ASR are observed. More recently, \cite{9893158} designed a DNN-based method to generate self-recoverable AEs that can be restored after blur, JPEG compression, crop, flip and resizing, by constructing coarse-grained adversarial perturbation with down and up sampling. Also, \cite{sun2022minimum} proposed to apply the generated AEs for privacy-preserving online photo sharing.

For achieving robustness against physical world degradations, \cite{athalye2018synthesizing} used C\&W in conjunction with proxy distributions to generate robust AEs against 2D transformations including rescaling, rotation, etc., and 3D transformations such as varying viewpoints and lighting conditions. In more specific applications, \cite{eykholt2018robust} proposed methods for generating robust physical adversarial perturbations for physical-world objects such as road signs against variant viewpoints of cameras. Along this line, \cite{wang2021exploring} improved the robustness of AEs against multiple affine transformations by leveraging the intermediate layer outputs of DNNs. \cite{tsai2020robust} proposed a novel method to generate robust adversarial 3D objects that preserved their properties after being physically manufactured by 3D printers.

\begin{center}
\begin{figure*}[t]
\centering{\includegraphics[width=0.8\textwidth]{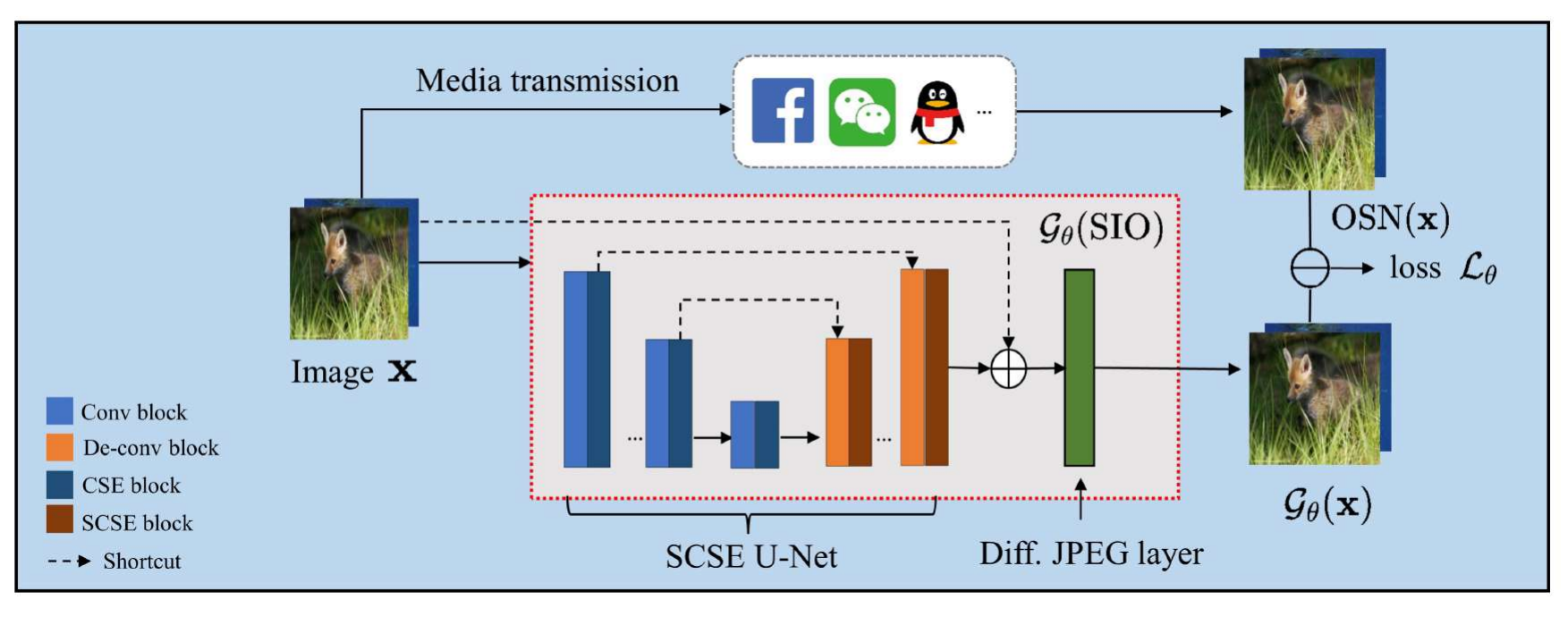}}
\caption{Our proposed SIO network for simulating the OSN image processing operations.}
\label{trainsio}
\end{figure*}
\end{center}

\subsection{Online Social Networks (OSNs)}
OSNs have been becoming dominant channels for acquiring, propagating, and storing images. There are some existing works on revealing the image processing operations in OSN platforms, by observing the differences between the uploaded and downloaded image pairs. Sun \emph{et al.} \cite{sunoptimal} pointed out that the image processing operations on Facebook mainly include format conversion, optional resizing, enhancement filtering, and JPEG compression. Specifically, the uploaded images are first transformed into the pixel domain where a truncation operation is used to ensure the pixel values are within [0,255]. Then if the resolution of the image exceeds 2048 pixels, this image will be resized. After that, the enhancement filtering is conducted for improving the appearance of the image. Due to its high adaptiveness, it is very challenging to accurately determine the enhancement filtering and the other postprocessing \cite{wei2016osn}. Finally, the image is compressed by the standard JPEG method where the $QF$ is dynamically determined by the quality and content of uploaded images. It should be noted that the $QF$ assignment mechanism is still mysterious, which makes the simulation for OSNs a nontrivial task.

There are some other works attempting to simulate partial OSN operations or behaviors through simple DNNs. For instance, Sun \textit{et al.} \cite{wei2020osn} predicted the modified image blocks (vulnerable blocks) resulting from Facebook transmission via a simple 8-layer neural network. Zhu \textit{et al.}\cite{zhu2022image} proposed two networks to imitate the scaling and enhancement filtering operations in OSNs.

\section{Deep Network for Simulating OSNs}\label{section:simosn}
Before diving into the details of generating robust AEs, let us first describe the simulation network for imitating the image processing procedures of OSNs. This is one of the key modules of the whole system, and the performance of such a simulation network directly affects the quality of the generated robust AEs. From the existing works \cite{wei2016osn, sunoptimal}, we know that OSN platforms manipulate transmitted images not only with JPEG compression but also with many unknown lossy manipulations. For the underlying AEs, the attack targets are mostly DNN models. Hence, the simulation network is expected to be differentiable such that it can be readily concatenated with the attack target. These considerations motivate us to design the \textbf{SI}mulation \textbf{O}SN (SIO) network, illustrated in Fig.\ref{trainsio}. Here, a JPEG layer is explicitly designed to reflect the fact that JPEG compression is ubiquitous in OSN platforms. The remaining encoder-decoder architecture then handles those unknown complicated operations in a unified manner. Note that our SIO is designed based on the in-depth understanding of the image processing procedures over OSNs. In contrast, some existing works, e.g., the ComModel \cite{wang2020towards}, used a simple CNN without specifically considering the nature of the OSN processing. As expected and will be verified experimentally, our SIO can much better imitate the behavior of OSNs and consequently produces more robust AEs.

More specifically, we adopt the ``Spatial Channel Squeeze-and-Excitation'' (SCSE) U-Net \cite{roy2018concurrent,roy2018recalibrating}, together with a shortcut design. Mathematically, denote SIO by $\mathcal{G}_{\theta}$, where $\theta$ represents its parameters. Upon receiving an image $\mathbf{x}$, the corresponding output of SIO, $\mathcal{G}_{\mathbf\theta}(\mathbf{x})$, is expected to be close to that of a given OSN platform. Now we are ready to present the details of the SIO network, namely, the SCSE U-Net module and differential JPEG layer.

\begin{center}
\begin{figure*}[t!]
\centering{\includegraphics[width=0.7\textwidth]{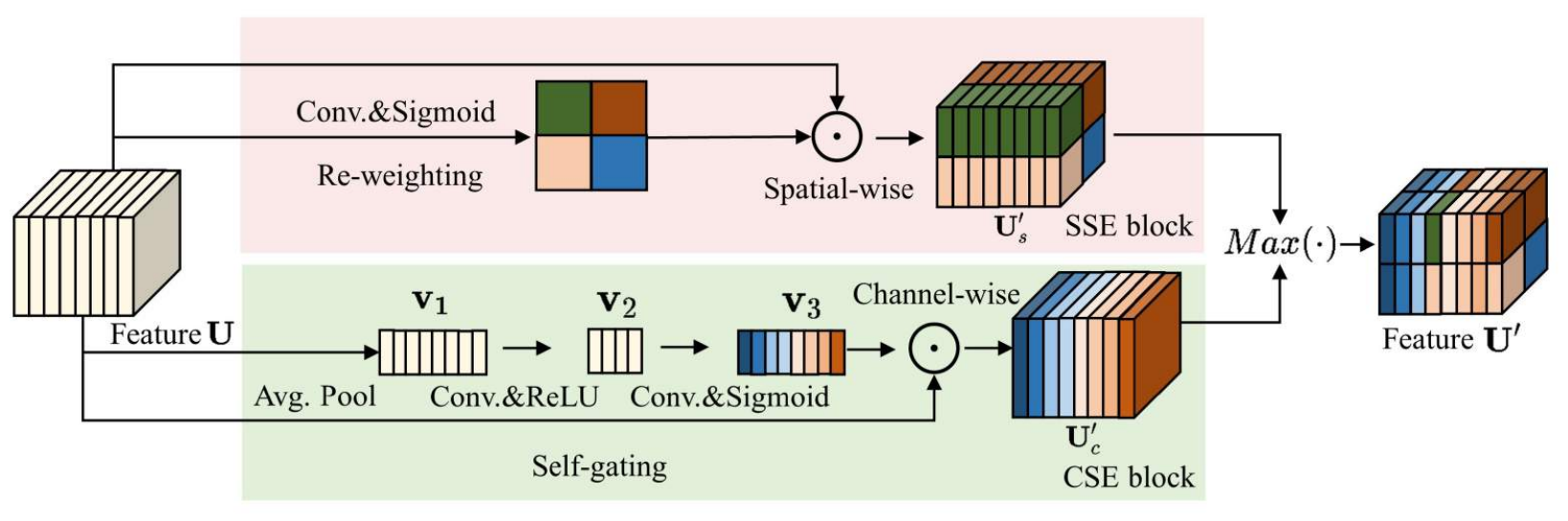}}
\caption{The illustration of the SCSE block.}
\label{fig:scse}
\end{figure*}
\end{center}
\subsection{SCSE U-Net Module}\label{section:scseunet}
Noticing that the OSN operation is essentially an image-to-image mapping, we adopt a U-Net as the backbone of the network, which is composed of four successive encoders and four symmetric decoders. Moreover, we append a shortcut for the residual learning from input images to the outputs of the U-Net, forcing it to pay more attention to the OSN noise generation, rather than the image content reconstruction.

To extract features of OSN degradation more efficiently and effectively, we specifically adopt the SCSE U-Net, denoted by $\Omega_{\theta}$, where the convolution blocks of the U-Net are concatenated with ``Channel Squeeze-and-Excitation''(CSE) blocks and deconvolution blocks are concatenated with SCSE blocks. Since SCSE U-Net has shown great potential in boosting the weight of meaningful features, and has been applied in semantic or medical image segmentation \cite{roy2018recalibrating,chatterjee2019building,kansal2019eyenet}, Magnetic Resonance Imaging (MRI) reconstruction \cite{huang2019mri}, image fusion\cite{ma2021sesf}, and skeletons extraction\cite{nathan2019skeletonnet}, we adopt it into our proposed SIO to imitate the image-to-image mapping of an OSN and extract the OSN operation features. The illustration of the SCSE block is given in Fig.\ref{fig:scse}.

Specifically, the SCSE block combines a ``Spatial Squeeze-and-Excitation'' (SSE) block in parallel to the CSE block, for concurrently recalibrating the input feature $\mathbf{U}\in\mathbb{R}^{H\times W\times C}$ both spatial-wise and channel-wise. The SSE block first re-weights $\mathbf{U}$ by the convolutional operator $\otimes$. Then the re-weighted feature is activated by a Sigmoid function and sequentially multiplied by $\mathbf{U}$ in a spatial-wise manner. Consequently, the spatial recalibrated feature $\mathbf{U}_s^\prime$ can be given as:

\begin{equation}\label{eq:sse}
\begin{gathered}
\mathbf{U}_s^\prime =\mathrm{Sigmoid}(\mathbf{W}_1\otimes\mathbf{U})\odot_s\mathbf{U},
\end{gathered}
\end{equation}
where $\mathbf{W}_1$ denotes a convolutional kernel and $\odot_s$ stands for the spatial-wise multiplication.
Switching to another branch of SCSE block, a global average pooling operation firstly squeezes feature $\mathbf{U}$ in channel-wise into an intermediate vector $\mathbf{v}_1\in\mathbb{R}^{1\times 1\times C}$. To fully capture channel-wise dependencies, $\mathbf{v}_1$ is enhanced to be the vector $\mathbf{v}_3$ by a self-gating operation, in which two fully connected layers $\mathbf{W}_2$ and $\mathbf{W}_3$ parameterize the gating mechanism to restrict model complexity and improve the generalizability. Finally, $\mathbf{U}$ will be re-scaled as $\mathbf{U}_c^\prime$ by the channel-wise multiplication with the channel activation $\mathbf{v}_3$. The pipeline of CSE block is formulated as
\begin{equation}\label{eq:cse}
\begin{gathered}
\mathbf{U}_c^\prime =\mathrm{Sigmoid}(\mathbf{W}_3\otimes\mathrm{ReLU}(\mathbf{W}_2\otimes\mathbf{v}_1))\odot_c \mathbf{U},
\end{gathered}
\end{equation}
where $\odot_c$ means the channel-wise multiplication.

Finally, we combine the CSE and SSE block to concurrently recalibrate the input $\mathbf{U}$ spatial-wise and channel-wise by a max-out operation to generate the re-weighted feature $\mathbf{U}^\prime$.

\begin{center}
\begin{figure*}[t!]
\centering{\includegraphics[width=0.6\textwidth]{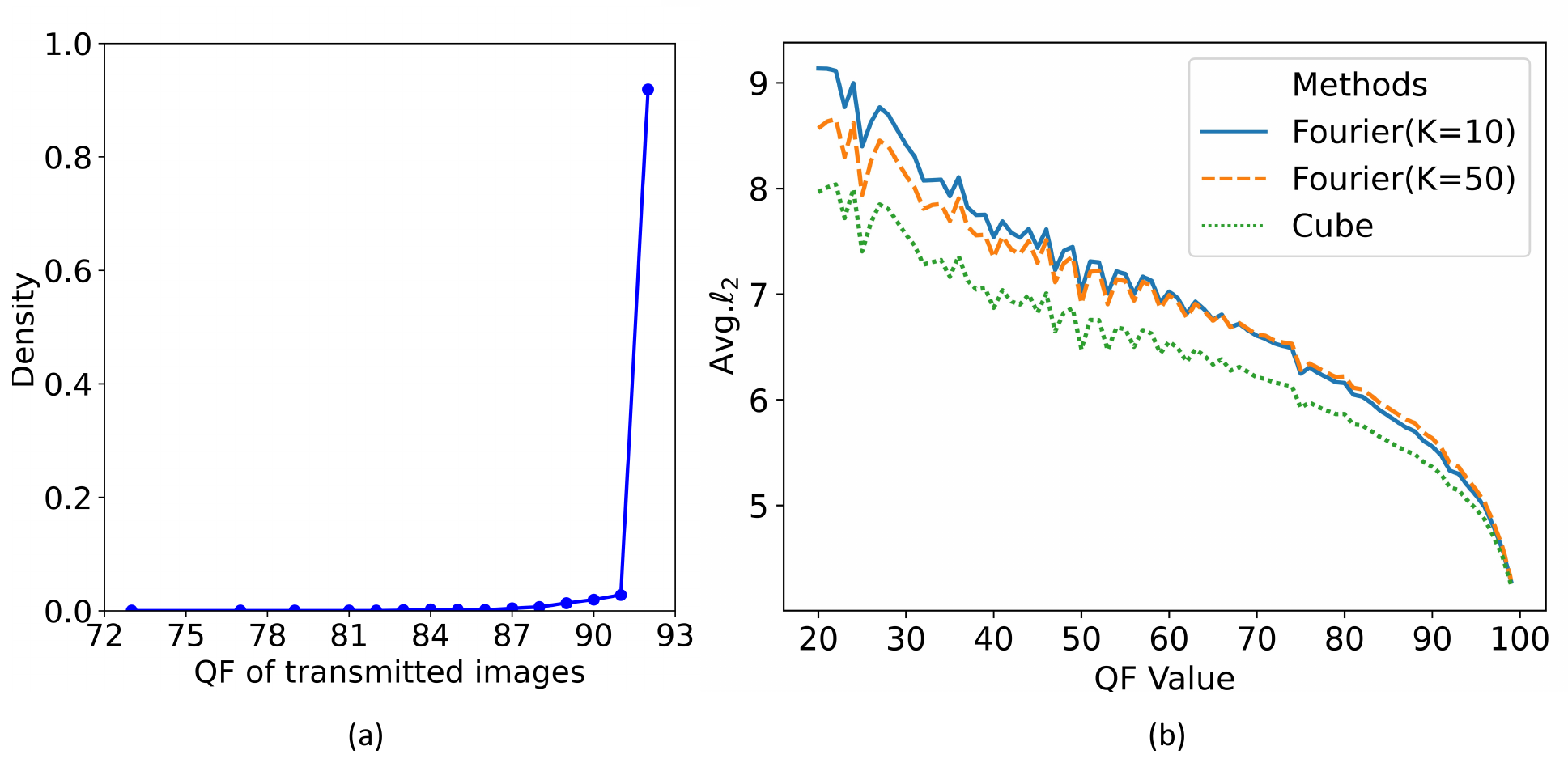}}
\caption{(a) The distribution of $QF$ values of images transmitted over Facebook. (b) Average $\ell_2$ errors of Cube \cite{shin2017jpeg} and Fourier \cite{xing2021invertible} with respect to different $QF$ values for approximating the JPEG method.}
\label{fig:jpegdifferror}
\end{figure*}
\end{center}

\subsection{Differentiable JPEG layer}\label{section:jpeg}
Another important ingredient of our proposed SIO network is a differentiable JPEG layer, which accounts for the JPEG compression ubiquitously integrated into almost all OSN platforms. Noticing that the encoding process in JPEG compression mainly consists of five steps: 1) color space conversion; 2) chroma subsampling; 3) block splitting; 4) discrete cosine transform (DCT) and 5) quantization by a preset $QF$ value, there are two essential problems to tackle for simulating a differentiable JPEG layer of a given OSN.

On the one hand, the JPEG layer requires a proper pre-defined $QF$ value for training and validation, which largely determines the simulation performance of the JPEG layer. We adopt a data-driven approach for exploring the $QF$ assignment strategy. Specifically, we randomly select 3000 clean images in \texttt{ImageNet} training set (JPEG format), resize them to $256\times256$, and then generate AEs in PNG format by randomly using traditional attack methods including FGSM, PGD, MIFGSM or C\&W. This allows us to take the influence of different attack methods on the $QF$ assignment into consideration. Later, these generated AEs are uploaded to the given OSN. In this way, we can then obtain the pairs of original AEs and corresponding OSN transmitted AEs. Here, $QFs$ of downloaded images are calculated by extracting the Quantization Table (QT) $Q$ from the JPEG header file and comparing it with the standard public QT $Q^0$. More details regarding this calculation can be referred to \cite{cozzolino2018noiseprint}. The distribution of images' $QFs$ after being transmitted over Facebook is depicted in Fig.\ref{fig:jpegdifferror} (a), from which we confirm that Facebook uses $QF=92$ to compress most uploaded images. Therefore, from the perspective of maximum likelihood, we assign the same $QF$ with the largest occurrence probability to all images for a given OSN. For example, we choose $QF=92$ for training SIO of Facebook.

On the other hand, the discontinuity of the rounding function $\left\lfloor\cdot\right\rceil$ in the quantization step makes it non-differentiable. In order to make the rounding operation differentiable, we can approximate the actual rounding function with a differentiable function by the Cube \cite{shin2017jpeg} or the Fourier methods \cite{xing2021invertible}. The former method introduces a polynomial as the rounding approximation, i.e.,

\begin{equation}\label{eq:cube}
\left\lfloor x \right\rceil_{c} = \left\lfloor x \right\rceil + (x - \left\lfloor x \right\rceil)^3,
\end{equation} where the maximum simulation error 0.125 arises upon rounding 0.5. The latter method proposes a Fourier series as an expansion of rounding function:
\begin{equation}\label{eq:Fourier}
\left\lfloor x \right\rceil_{f}(K) = x - \frac{1}{\pi}\sum\nolimits^{K}_{k=1}\frac{(-1)^{k+1}}{k} sin(2\pi kx),
\end{equation}
where the parameter $K$ controls the approximation accuracy. The larger $K$ would correspond to smaller simulation errors between (\ref{eq:Fourier}) and the actual rounding function \cite{xing2021invertible}. To compare these two types of approximation, we compute the average approximation error over 1000 images from \texttt{ImageNet}. As shown in Fig.\ref{fig:jpegdifferror} (b), the Cube approximation consistently leads to the lowest $\ell_2$ error for all $QFs$. Hence, we adopt it in our differentiable JPEG layer. As for the other steps in JPEG procedures, they are easy to be implemented by differentiable functions. Eventually, we obtain a differentiable approximation layer of JPEG denoted by $\hat{\mathcal{J}}$.

\subsection{The training loss}\label{section:trainloss}

For training SIO, we need to collect pairs of input and output of an OSN platform. More specifically, the purpose of designing the simulation network SIO is to assist the subsequent robust AE generation. Hence, for training the SIO for Facebook, we generate 3000 AEs based on clean images chosen randomly from the training set of \texttt{ImageNet} against ResNet-50 as the inputs, upload them to the given OSN, and download the OSN processed images as outputs. Here, although we only utilize a single target model ResNet-50 to generate AEs, we observe that the SIO trained in this manner performs satisfactorily on other target models with different architectures as well. Such a phenomenon may be because SIO, as a simulator for the overall OSN image processing operations, does not rely heavily on the type of AE inputs; but rather, more on the correspondence of the input-output pairs. Overall, the training procedure is illustrated in Fig.\ref{trainsio}, where the training objective for minimizing loss $\mathcal{L}_\theta$ can be defined as follows:

\begin{equation}\label{eq:mseloss}
\begin{gathered}
\mathop{\min}\limits_{\mathbf\theta}{\frac{1}{M}\sum\nolimits_{i=0}^{M-1}\mathcal{D}( \mathrm{OSN}(\mathbf{x}_{i}),\mathcal{G}_{\mathbf\theta}(\mathbf{x}_i))},
\end{gathered}
\end{equation} where $\mathrm{OSN}(\mathbf{x}_i)$ is the transmitted version of the $\mathbf{x}_i$ from a the given OSN, $M$ denotes the number of training samples, and $\mathcal{D}(\cdot,\cdot)$ represents a distance metric. For simplicity, we use $\ell_2$ Euclidean distance in this paper. Noticing the SCSE U-Net module $\Omega_{\theta}$ and the JPEG layer $\hat{\mathcal{J}}$, the above objective function can be further written as

\begin{equation}\label{eq:mseloss3}
\begin{gathered}
\mathop{\min}\limits_{\mathbf\theta}{\frac{1}{M}\sum\nolimits_{i=0}^{M-1}\Big \|  \mathrm{OSN}(\mathbf{x}_{i})-\hat{\mathcal{J}}\Big(\Omega_{\theta}(\mathbf{x}_i)+\mathbf{x}_i,q\Big)\Big \|_2},
\end{gathered}
\end{equation} where the parameter $q$ denotes the $QF$ employed. For Facebook/WeChat/QQ, we set $q=92/58/85$ according to the data-driven approach aforementioned in Section \ref{section:jpeg}.

\begin{center}
\begin{figure*}[t]
\centering{\includegraphics[width=0.9\textwidth]{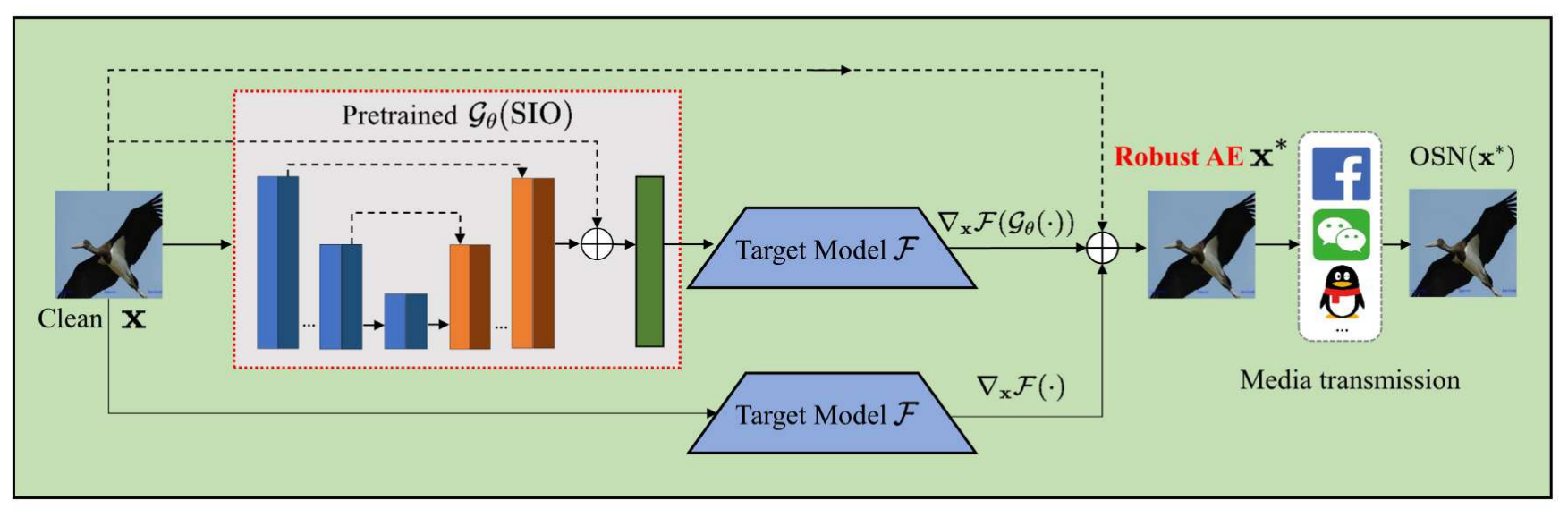}}
\caption{An illustration of our proposed attack framework for generating robust AEs.}
\label{genae}
\end{figure*}
\end{center}
\section{Generation of robust AE against OSNs}\label{section:genae}
Upon having the well-trained SIO network $\mathcal{G}_\theta$, we now discuss the problem of generating robust AEs against the OSN transmission. The scheme is demonstrated in Fig.\ref{genae}, where $\mathcal{G}_\theta$ is concatenated with the target model $\mathcal{F}$. Note that we in this work only consider the white-box attack, and hence, $\mathcal{F}$ is assumed to be available. Similar to the presentation of many attack methods, without losing the generality, we further assume that $\mathcal{F}$ is a classification model.

More specifically, for an image classifier $\mathcal{F}(\mathbf{x}): \mathbf{x}\in\mathbb{R}^d\rightarrow \hat{y}\in \left \{ 0,1,...,c-1 \right \}$ with $c$ classes, the input is a $d$ dimensional image and the output is a scalar $\hat{y}$ as the prediction label. When presenting our proposed attack method, we confine ourselves to the non-targeted attack, and the extension to the targeted attack shall be straightforward. The goal of the attacker is to craft an adversarial image $\mathbf{x}^*$ in the vicinity of the normal $\mathbf{x}$ that deceives the classifier into predicting a label not equal to the ground truth $y$. Also, we are only interested in those $\mathbf{x}$ whose classification result $\hat{y}$ is correct, namely, $y=\hat{y}$.

Clearly, for a successful AE $\mathbf{x}^*$ with respect to $\mathcal{F}$, we have $\mathcal{F}(\mathbf{x}^*)\neq{y}$. Also, $\mathbf{x}^*$ is expected to be robust against OSN transmission, and hence, $\mathcal{F}(\hat{\mathbf{x}})\neq{y}$ shall hold as well, where $\hat{\mathbf{x}}=\mathrm{OSN}(\mathbf{x}^*)$. Noticing the fact that the generated AEs may or may not be actually transmitted over OSNs, we should maintain the attack capabilities of both $\mathbf{x}^*$ (without OSN transmission) and $\hat{\mathbf{x}}$ (with OSN transmission). Therefore, we formulate the following optimization framework for generating the robust AEs $\mathbf{x}^*$:

\begin{equation}\label{eq:osnuntattack}
\begin{gathered}
\mathop{\min}\limits_{\mathbf{x}^*} \ \mathcal{D}(\mathbf{x},\mathbf{x}^*),\\
s.t. \ \mathcal{F}(\mathbf{x}^*)\neq{y},\\
\mathcal{F}(\hat{\mathbf{x}})\neq{y},\\
 \mathbf{x}^*\in[0, 1]^n,
\end{gathered}
\end{equation} where the metric $\mathcal{D}$ measures the distance between the AE $\mathbf{x}^*$ and the normal image $\mathbf{x}$, and could be set as $\ell_0$, $\ell_2$ and $\ell_\infty$ norm in the Euclidean space. In addition, the last constraint ensures that the resulting $\mathbf{x}^*$ is a valid image.

In the following, we derive the Lagrange-form solutions and the gradient projection solutions, which are analogous to the two mainstreams of traditional adversarial attack methods.

\subsection{Lagrange-form solution}\label{section:cw}
Due to the highly non-linear inequality constraints, it is difficult to solve (\ref{eq:osnuntattack}) directly. We now use a technique, motivated by the one introduced in C\&W \cite{carlini2017towards}, to solve the optimization problem (\ref{eq:osnuntattack}). The Lagrange-form solution to be derived below is termed as R-C\&W, where `R-' stands for a robust version. We first define a function
\begin{equation}\label{eq:cw1}
\begin{gathered}
f(\mathbf{x}^*,y,\mathcal{Z}) 
= \mathop{\max}\left\{ \mathcal{Z}(\mathbf{x}^{*})[y]-\mathop{\max}\limits_{i\neq y}\mathcal{Z}(\mathbf{x}^{*})[i],  -k \right\},
\end{gathered}
\end{equation}
such that $\mathcal{F}(\mathbf{x}^*)\neq{y}$ if and only if $f(\mathbf{x}^*,y,\mathcal{Z})\leq0$, where $k$ is an adjustable parameter to control the confidence for finding an AE, $\mathcal{Z}(\cdot)=[z_0,z_1,...,z_{c-1}]$ represents the logits output vector of the penultimate layer in the target model $\mathcal{F}$, and $\mathcal{Z}(\mathbf{x}^*)[i]$ stands for the $i$th element of $\mathcal{Z}(\mathbf{x}^*)$. With the above function $f$ at hand, the problem (\ref{eq:osnuntattack}) can be re-formulated as

\begin{equation}\label{eq:osnuntattack2}
\begin{gathered}
\mathop{\min}\limits_{\mathbf{x}^*} \ \mathcal{D}(\mathbf{x},\mathbf{x}^*),\\
s.t.~ \ f(\mathbf{x}^*,y,\mathcal{Z})\leq0,\\
 f(\hat{\mathbf{x}},y,\mathcal{Z})\leq0,\\
 \mathbf{x}^*\in[0, 1]^n.
\end{gathered}
\end{equation}

To deal with the box constraint $\mathbf{x}^*\in[0, 1]^n$, we introduce a new variable $\mathbf{w}^{*}\in\mathbb{R}^d$ and define a function $h$ as

\begin{equation}\label{eq:cw22}
\begin{gathered}
h(\mathbf{w}^{*}) = (1+\mathop{\tanh}(\mathbf{w}^{*}))/2.
\end{gathered}
\end{equation} Noting that $-1\leq \tanh(\mathbf{w}^{*}) \leq 1$, $h(\mathbf{w}^{*})$ will be guaranteed to fall into the interval $[0,1]$. This allows us to remove the box constraint by replacing $\mathbf{x}^*$ with $h(\mathbf{w}^{*})$. Then (\ref{eq:osnuntattack2}) can be expressed as:

\begin{equation}\label{eq:osnuntattack3}
\begin{gathered}
\mathop{\min}\limits_{\mathbf{w}^{*}} \ \mathcal{D}(\mathbf{x},h(\mathbf{w}^{*})),\\
s.t. ~\ f(h(\mathbf{w}^{*}),y,\mathcal{Z})\leq0,\\
 f(\mathcal{G}_\theta(h(\mathbf{w}^{*})),y,\mathcal{Z})\leq0.\\
\end{gathered}
\end{equation}

\begin{algorithm}[!t]
	\begin{small}
	\def\dx{0.5in}
		\caption{The $\ell_2$ constrained non-targeted attack algorithm for generating robust AEs by R-C\&W.}
		\KwIn{Logits output $\mathcal{Z}$ of the attack target model; a pre-trained OSN simulator $\mathcal{G}_\theta$; a clean image $\mathbf{x}$, the ground truth label $y$; constants $c>0, \lambda\in(0,1)$; the iterative steps $T>0$.}
		\KwOut{Adversarial image $\mathbf{x}^*$}
		\mbox{Initialize $\ell_2^{*} = \mathbf{e}^{10}$, $\mathbf{w}_0 = \frac{1}{2}\ln{\frac{\mathbf{x}}{1-\mathbf{x}}}$}  \hspace*{\fill} \mbox{$\triangleright$ Project $\mathbf{x}$ to tanh space} \\
		\For {$t = 0$ to $T-1$}{
			
\makebox{\shortstack[l]{Update $\mathcal{L}_{adv}^{t}(\mathbf{w}_t,\mathbf{w},y,\mathcal{Z},\mathcal{G}_\theta)=c \cdot \mathcal{D}(h(\mathbf{w}_t),h(\mathbf{w}))$ \\ + $\lambda \cdot f(h(\mathbf{w}_t),{y},\mathcal{Z})+ (1-\lambda)\cdot f(\mathcal{G}_\theta(h(\mathbf{w}_t)),{y},\mathcal{Z})$}}
\hspace*{\fill} \mbox{$\triangleright$ Eq. (\ref{eq:cw3})}\\
			\mbox{Obtain gradient $\mathbf{g}_t = \nabla_{\mathbf{w}_t} \mathcal{L}_{adv}^{t} $}  \\
			\mbox{Update $\mathbf{w}_t$ to $\mathbf{w}_{t+1}$ by Adam optimizer and $\mathbf{g}_t$}\\
			\mbox{$\mathbf{x}_{t+1} = \frac{1}{2}(\tanh{(\mathbf{w}_{t+1})}+1)$} \mbox{$\triangleright$ Eq. (\ref{eq:cw22})}\\
			\uIf{$\parallel\mathbf{x}_{t+1} - \mathbf{x}\parallel_{2} < \ell_2^{*}$ and $\mathcal{F}(\mathbf{x}_{t+1}) \neq y$}{
			  $\ell_2^{*} = \parallel\mathbf{x}_{t+1} - \mathbf{x}\parallel_{2} $\;
			  $\mathbf{x}^{*} = \mathbf{x}_{t+1}$\;  \hspace*{\fill} \mbox{$\triangleright$ Update adversarial image $\mathbf{x}^{*}$}
			}
			\uIf{$\mathop{\max}\left\{t\mod \lfloor \frac{T}{10} \rfloor == 0,1\right\}$ and $\mathcal{L}_{adv}^{t} > \mathcal{L}_{adv}^{t-1}$}{
			\Return $\mathbf{x}^{*}$.  \hspace*{\fill} \mbox{$\triangleright$ Early stop when the loss does not converge}
			}
			
		}		
		\Return $\mathbf{x}^{*}$. \\
	\end{small}
\end{algorithm}

Further, this inequality-constrained optimization problem can be written into an unconstrained problem by using Lagrange multipliers. We then have
\begin{equation}\label{eq:cw2}
\begin{gathered}
\mathop{\min}\limits_{\mathbf{w}^*}\mathcal{L}_{adv}(\mathbf{w}^{*},\mathbf{w},y,\mathcal{Z},\mathcal{G}_\theta)\\
\end{gathered}
\end{equation}
where
\begin{equation}\label{eq:cw3}
\begin{gathered}
\mathcal{L}_{adv}(\mathbf{w}^{*},\mathbf{w},y,\mathcal{Z},\mathcal{G}_\theta)=c \cdot \mathcal{D}(h(\mathbf{w}^{*}),h(\mathbf{w}))\\
+\lambda \cdot f(h(\mathbf{w}^{*}),{y},\mathcal{Z})
+ (1-\lambda)\cdot f(\mathcal{G}_\theta(h(\mathbf{w}^{*})),{y},\mathcal{Z}) .
\end{gathered}
\end{equation}
Here, the parameter $c>0$ regulates the equilibrium between the fidelity term and the two misclassification terms. Also, $\lambda\in(0,1)$ trades off the relative importance of the attack capabilities of the AEs before and after the OSN transmission. Specifically, if $\lambda$ is large, the second term dominates $\mathcal{L}_{adv}$ in (\ref{eq:cw3}), resulting in AEs with greater attack capabilities against $\mathcal{F}$ than $\mathcal{F}(\mathcal{G_\theta}(\cdot))$. Conversely, with a small $\lambda$, the dominance of the third term exceeds that of the second term, which is more beneficial for the robust AE generation against $\mathcal{F}(\mathcal{G_\theta}(\cdot))$. A more specific analysis regarding the influence of $\lambda$ on ASR will be presented in the next section.

Back to the unconstrained optimization problem (\ref{eq:cw2}), it can be easily solved by resorting to some existing solvers. In this work, we adopt the Adam\cite{kingma2014adam} optimizer to update $\mathbf{w}^{*}$, since it converges substantially more quickly than other solvers such as standard gradient descent and gradient descent with momentum\cite{carlini2017towards}. The detailed procedure of generating robust AE via the proposed R-C\&W is given in Algorithm 1. More specifically, as shown in lines 7-8 of Algorithm 1, we renew $\mathbf{x}^{*}$ by $\mathbf{x}_{t+1}$ at the iteration $t$ if 1) the updated $\ell_2$ perturbation is smaller than that in the previous iterations and 2) $\mathbf{x}_{t+1}$ successfully fools the target model $\mathcal{F}$. Condition 2) makes sure that the generated AEs attack $\mathcal{F}$ successfully.  We do not force AEs to successfully attack $\mathcal{F}(\mathcal{G_\theta}(\cdot))$ at the same time in 2) for improving update frequencies. In addition, we find that a small $\lambda$ always ensures the attack ability against $\mathcal{F}(\mathcal{G_\theta}(\cdot))$ in practice. Moreover, condition 2) can be flexibly adjustable. If generated AEs are more likely to be transmitted over OSNs, the condition 2) could be set as $\mathcal{F}(\mathcal{G_\theta}(\mathbf{x}_{t+1}))\neq y$ instead of $\mathcal{F}(\mathbf{x}_{t+1}) \neq y$, so as to more focus on the successful attack after being transmitted over an OSN.

From (\ref{eq:osnuntattack}) to (\ref{eq:cw3}), we can see that it is challenging to have a determined bound of the perturbation magnitude, as the fidelity term is included in the optimization objective. However, in some applications, it is highly desirable to explicitly have a bound for the perturbation magnitude. This motivates us to develop the following gradient projection solutions to the robust AE generation problem in (\ref{eq:osnuntattack}).

\begin{algorithm}[!t]
	\begin{small}
		\caption{The gradient projection non-targeted attack algorithm for generating robust AEs by R-MIFGSM.}
		\KwIn{Logits output $\mathcal{Z}$ of the attack target model; a pre-trained OSN simulator $\mathcal{G}_\theta$; a clean image $\mathbf{x}$; the iterative step $T>0$; the distortion budget $\epsilon$; the ground truth label $y$; the decay factor $\mu$.}
		\KwOut{Adversarial image $\mathbf{x}^*$ with $\parallel\mathbf{x}^*-\mathbf{x}\parallel_\infty \le\epsilon$}
		Initialize \mbox{$\mathbf{x}_{0}=\mathbf{x}, \mathbf{g}_0=0,\alpha=\epsilon/T$} \mbox{$\triangleright$ \cite{dong2018boosting}}\\
		\For {$t = 1$ to $T$}{
			\mbox{$\mathbf{g}_{t} = \mu\cdot\mathbf{g}_{t-1}+\frac{\nabla_{\mathbf{x}_{t-1}}\mathcal{C}_{adv}(\mathbf{x}_{t-1},y,\mathcal{Z},\mathcal{G}_\theta)}{\parallel \mathcal{C}_{adv}(\mathbf{x}_{t-1},y,\mathcal{Z},\mathcal{G}_\theta)\parallel_1}$} \hspace*{\fill}\\
			\mbox{$\mathbf{x}_{t} = \mathbf{x}_{t-1}+ \lvert\alpha\rvert \cdot sign(\mathbf{g}_{t})$} \\
			\mbox{$\mathbf{\delta}_t = \mathop{clip}(\mathbf{x}_{t}-\mathbf{x},-{\epsilon},+{\epsilon})$}\\
			\mbox{$\mathbf{x}_{t} = \mathop{clip}(\mathbf{x}_{t-1}+\mathbf{\delta}_t ,0,1)$}\\
		}		
		\Return $\mathbf{x}^{*} = \mathbf{x}_{T}$. \\
	\end{small}
\end{algorithm}

\subsection{Gradient projection solutions}
We now present an alternative, gradient projection solution, to the problem (\ref{eq:osnuntattack}). This is analogous to the series of traditional works including FGSM, PGD, and MIFGSM, for generating AEs. Noticing that the inequality constraints in (\ref{eq:osnuntattack}) make the optimization problem challenging, we propose to replace them with simpler forms. A viable solution is to use the cross-entropy (CE) loss to measure the performance of the classifier $\mathcal{F}$: the larger the CE loss, the higher the possibility of misleading the target model. As a result, the constraint $\mathcal{F}(\mathbf{x}^*)\neq{y}$ can be replaced by $\mathrm{CE}(\mathcal{Z}(\mathbf{x}^*),{y})\geq a$, where $a$ is a threshold. Then the problem (\ref{eq:osnuntattack}) can be recast into

\begin{equation}\label{eq:osnuntattackg}
\begin{gathered}
\mathop{\min}\limits_{\mathbf{x}^*} \ \mathcal{D}(\mathbf{x},\mathbf{x}^*),\\
s.t. \ \mathrm{CE}(\mathcal{Z}(\mathbf{x}^*),{y})\geq a,\\
\mathrm{CE}(\mathcal{Z}(\hat{\mathbf{x}}),{y})\geq b,\\
 \mathbf{x}^*\in[0, 1]^n,
\end{gathered}
\end{equation} where $b$ is another parameter.

Similar to FGSM, PGD, and MIFGSM, a simple approximation of the solution to (\ref{eq:osnuntattackg}) is to maximize CE loss terms for exceeding the thresholds, and then pull back the perturbed $\mathbf{x}^*$ to an allowed vicinity of $\mathbf{x}$. The optimization objective now becomes

\begin{equation}\label{eq:jointce}
\begin{gathered}
\mathop{\max}\limits_{\mathbf{x}^*}\mathcal{C}_{adv}(\mathbf{x}^*,y,\mathcal{Z},\mathcal{G}_\theta)\\
\end{gathered}
\end{equation} where

\begin{equation}\label{eq:jointce2}
\begin{gathered}
\mathcal{C}_{adv}(\mathbf{x}^*,y,\mathcal{Z},\mathcal{G}_\theta)= \\
\lambda \cdot \mathrm{CE}(\mathcal{Z}(\mathbf{x}^*),{y})+(1-\lambda) \cdot \mathrm{CE}(\mathcal{Z}(\mathcal{G}_\theta(\mathbf{x}^*)),{y}).
\end{gathered}
\end{equation} Here, the parameter $\lambda$ balances the attack capabilities of AEs before and after the OSN transmission.

Below, we present three types of robust AE generation solutions, namely, R-FGSM, R-PGD, R-MIFGSM to (\ref{eq:jointce}), corresponding to the traditional attack methods FGSM, PGD, and MIFGSM, respectively. 

For R-FGSM, by linearizing the cost function in (\ref{eq:jointce2}), we can obtain an optimal max-norm constrained perturbation $\lvert\epsilon\rvert\cdot\mathop{sign}(\nabla_{\mathbf{x}}\mathcal{C}_{adv})$, where $\epsilon$ can be regarded as the radius of $l_\infty$ ball with $\mathbf{x}$ being the center. In this case, the robust AEs can be derived as

\begin{equation}\label{eq:fgsmosn}
\begin{gathered}
\mathbf{x}^*= \mathop{clip}(\mathbf{x} + \lvert\epsilon\rvert\cdot\mathop{sign}(\nabla_{\mathbf{x}}\mathcal{C}_{adv}),0,1),
\end{gathered}
\end{equation}
where a clip function guarantees the result to be in the interval of $[0,1]$.

A powerful multi-step variant of R-FGSM is R-PGD, which substantially performs projection descent on the negative $\mathcal{C}_{adv}$ loss as
\begin{equation}\label{eq:pgd}
\begin{gathered}
\begin{cases}
\mathbf{x}_{0} = \mathbf{x},\\
\mathbf{g}_{t} = \nabla_{\mathbf{x}_{t-1}}\mathcal{C}_{adv}(\mathbf{x}_{t-1},y,\mathcal{Z},\mathcal{G}_\theta),\\
\mathbf{x}_{t}=\prod\limits_{\mathcal S}(\mathbf{x}_{t-1} + \lvert\alpha\rvert \cdot sign( \mathbf{g}_{t})),
\end{cases}
\end{gathered}
\end{equation}
where $\alpha$ denotes the step size and $\mathbf{g}_{t}$ is the gradient at iteration $t$. Here, $\prod$ represents the projection operation such that the result is within a feasible region $\mathcal{S}=\left\{\mathbf{x}^{\prime}|\mathcal{D}(\mathbf{x},\mathbf{x}^{\prime})\leq\epsilon\right\}$ at each iteration.  R-PGD will stop when $t$ reaches a preset value, and then the generated AE becomes $\mathbf{x}^*=\mathbf{x}_{t}$.

For R-MIFGSM, it accumulates a momentum term with a decay factor $\mu$ in the gradient direction of the loss function across iterations. The detailed procedure is given in Algorithm 2, as a representative for this stream of gradient projection solutions.

\begin{table}[t]
  \centering
  \setcounter{table}{0} 
  \caption{Attack parameter settings to generate AEs for training.}
  \scalebox{0.7}{
    \begin{tabular}{l|l}
    \hline
    \hline
    Methods & Parameters \Tstrut\\
    \hline
    FGSM  & $\epsilon$=3. \Tstrut\\
    \hline
    PGD   & $\epsilon$=3; $\alpha$=2/255; $T$=40. \Tstrut\\
    \hline
    MIFGSM & $\epsilon$=3; $\alpha$=$\epsilon/T$; $T$=5; $\mu$=1.\Tstrut \\
    \hline
    C\&W    & $c$=1; $k$=0; $T$=40. \Tstrut\\
    \hline
    \hline
    \end{tabular}%
    }
  \label{tab:attpara}%
\end{table}%

So far, we have introduced four novel AE generation methods R-C\&W, R-FGSM, R-PGD, and R-MIFGSM based on our SIO network. It should be pointed out that our attack framework can essentially accommodate any optimization solutions relying on the gradient information.

\section{Experimental Results}\label{section:expe}
In this section, we present the experimental results of our proposed method and the comparison with the state-of-the-art competitors. To begin with, we give the experimental settings. Then, we demonstrate the robustness of AEs transmitted over Facebook, and report the superiority of our method over the competing approaches. In this work, we mainly use Facebook as a representative OSN to show the performance, and also validate that our proposed methods can be readily extended to other OSN platforms such as WeChat and QQ.

\subsection{Experimental Setup}\label{section:exset}

\subsubsection{Datasets}
The training set of \texttt{ImageNet} is used for generating AEs as the training dataset of the SIO or ComModel\cite{wang2020towards} for simulating OSNs. Then we evaluate the OSN simulation and attack performance both in the validation sets of \texttt{ImageNet} and \texttt{Cifar-10}. Note that there is no overlap between the training and validation sets.
\subsubsection{Attack settings}
For training the SIO, we generate AEs based on clean images randomly chosen from the \texttt{ImageNet} training set, by vanilla FGSM, MIFGSM, PGD or C\&W against ResNet-50 model pre-trained on \texttt{ImageNet}. The training attack parameters are listed in Table \ref{tab:attpara}, which are all typical settings and selected according to the following criteria: 1) they are in line with the optimal settings suggested in the respective original papers and widely-adopted attack libraries such as cleverhans\footnote{https://github.com/cleverhans-lab/cleverhans.} and torchattacks\footnote{https://github.com/Harry24k/adversarial-attacks-pytorch.}; 2) they ensure the imperceptibility of the generated adversarial perturbations; and 3) the attack success rate is high enough so as to effectively generate a large number of AEs for training. Additionally, these parameters do affect both the simulation and robustness performance; but their impacts are negligible. More details can be found in Appendix \ref{sec:appa2}. For testing, the parameters of gradient projection solutions are the same as those shown in Table \ref{tab:attpara} except for $\epsilon$, which will be specified in the following subsections. As for the Larange-form solutions, the parameter $k$ in (\ref{eq:cw1}) is set to be a constant 0, and all the remaining ones are set in a way that the average $l_2$ values of the generated AEs are similar.     


\subsubsection{Training details}
For training the SIO, we adopt the Adam \cite{kingma2014adam} optimizer with $\beta_1 = 0.9$ and $\beta_2=0.999$. The learning rate is initialized as $1e-4$, and will be reduced by half if the loss value defined in (\ref{eq:mseloss3}) fails to decrease for 10 epochs until the convergence. For a fair comparison, we retrain the ComModel \cite{wang2020towards} for Facebook, WeChat or QQ with their public training parameters and the same training dataset as those of our SIO model. The training loss and image pairs have been introduced in Section \ref{section:trainloss}.

\subsubsection{Evaluation criteria}
We define the following two metrics to evaluate the performance of different AE generation methods, from the perspective of the attack capabilities before and after the OSN transmissions. Let $N$ be the total number of perturbed images generated by a certain attack method against the target model $\mathcal{F}$. Among these $N$ perturbed images, there are $N_1$ and $N_2$ images that can successfully fool $\mathcal{F}$ before and after the OSN transmission, respectively. More specifically, $N_1$ and $N_2$ can be represented as follows:
 
\begin{equation}\label{eq:n1n2}
\begin{gathered}
N_1 = \sum\nolimits_{i=1}^{N}\mathbb{I}(\mathcal{F}(\tilde{\mathbf{x}}_i)\neq{y_i}), \ 
N_2 = \sum\nolimits_{i=1}^{N}\mathbb{I}(\mathcal{F}(\mathrm{OSN}(\tilde{\mathbf{x}}_i))\neq{y_i}),\\
\end{gathered}
\end{equation}
where $\tilde{\mathbf{x}}_i$ represents the $i$-th perturbed image, $y_i$ is its ground truth label, and $\mathbb{I}$ means the indicator function that yields $1$ when the input is true and $0$ otherwise. Then we can define the following two types of attack success rate (ASR):

\begin{equation}\label{eq:asr}
\begin{gathered}
ASR = N_1/N ~\mathrm{and}~ ASR' = N_2/N,
\end{gathered}
\end{equation} which measure the attack capabilities of AEs before and after the OSN transmission. The number of test images on Facebook, WeChat, and QQ are 200, 100, and 50, respectively. The relatively small number of test images is due to the constraints imposed by the OSNs on the number of uploaded images. The category distributions of test sets for different OSNs are uniform, reducing the category bias of test sets.

\subsubsection{Comparative methods}
We compare our proposed methods with two state-of-the-art competing schemes ComModel\cite{wang2020towards} and \textbf{JPEG}-\textbf{R}esistant (JPEGR) \cite{shin2017jpeg}. To the best of our knowledge, ComModel is the only work aiming at generating robust AEs over OSNs so far. Also, JPEGR was reported to show competitive performance with ComModel in \cite{wang2020towards} when resisting JPEG compression.

\begin{table*}[t!]
\setlength\tabcolsep{3pt}
  \centering
  \setcounter{table}{1} 
  \caption{Robustness performance ($\%$) of AEs generated by R-C\&W (ours), C\&W integrated with ComModel\cite{wang2020towards}, JPEGR\cite{shin2017jpeg} and vanilla C\&W. For fairness, the absolute difference values of Avg. $\ell_2$ among compared methods are no more than 0.1.}

  \scalebox{0.65}{
    \begin{tabular}{c|ccc|ccc|ccc|ccc|ccc|ccc}
    \hline
    \hline
    \multirow{4}{*}{Methods}    & \multicolumn{12}{c|}{Facebook}&\multicolumn{3}{c|}{WeChat}&\multicolumn{3}{c}{QQ}\Tstrut\\
   \cline{2-19}     & \multicolumn{9}{c|}{ImageNet}&\multicolumn{3}{c|}{Cifar-10}& \multicolumn{3}{c|}{ImageNet}& \multicolumn{3}{c}{ImageNet}\Tstrut\\
  \cline{2-19}     & \multicolumn{3}{c|}{VGG-19bn}  & \multicolumn{3}{c|}{ResNet-50}   & \multicolumn{3}{c|}{Inception-V3} &\multicolumn{3}{c|}{ResNet-18}&    \multicolumn{3}{c|}{VGG-19bn}&  \multicolumn{3}{c}{VGG-19bn}\Tstrut\\
\cline{2-19}          &  ASR  & ASR$^\prime$  & Avg.$\ell_2$&  ASR  & ASR$^\prime$& Avg.$\ell_2$   & ASR  & ASR$^\prime$  &  Avg.$\ell_2$& ASR  & ASR$^\prime$  &  Avg.$\ell_2$& ASR  & ASR$^\prime$  & Avg.$\ell_2$& ASR & ASR$^\prime$  &  Avg.$\ell_2$\Tstrut  \\
    \hline
    R-C\&W (Ours)    &   \textbf{100}  & \textbf{70.0}    & 0.601&\textbf{100}  & \textbf{70.5} & 0.747& \textbf{100}  &  \textbf{88.0}  &  0.741& \textbf{100}    & \textbf{63.5}  & 0.427 &\textbf{100}&\textbf{65.0}&2.942& \textbf{100}    & \textbf{60.0}    &2.227\Tstrut \\
     \hline
         C\&W w/ ComModel &  31.0  & 21.5  & 0.602& 43.0  & 30.5   & 0.798& 27.0  & 39.5    & 0.733  & 49.5  & 35.5  & 0.433 & 39.0    & 48.0   &2.969& 79.0    & 54.0   &2.389\Tstrut \\

   \hline
    C\&W w/ JPEGR &  14.5  & 65.0  & 0.597& 24.5  & 64.0   & 0.777& 20.5  & 79.0    & 0.745  & 14.5  & 28.5  & 0.425 & 35.0   & 47.0   &2.984& 60.0    & 43.0    & 2.394\Tstrut \\
   \hline
    Vanilla C\&W  &\textbf{100}  & 12.0   & 0.591& \textbf{100} & 19.0    & 0.784 &   \textbf{100}  & 26.5  & 0.742& \textbf{100}   & 27.5  & 0.423 & \textbf{100}    & 22.0    &2.953 & \textbf{100}   & 20.0    &2.305\Tstrut  \\
    \hline
    \hline
    \end{tabular}
    }%
  \label{tab:cwtotal}%
\end{table*}%

\subsection{Robustness evaluations}\label{section:comp1}
We now demonstrate the ASR performance of the AEs generated by comparative methods before and after the OSN transmission. Here, we mainly focus on the OSN platform Facebook, when evaluating the performance of attacking multiple models as well as different datasets. Due to the space limit and similar observations, we also report the performance comparison on WeChat and QQ, by fixing one specific model and one dataset. We first give the results of the $\ell_2$ constrained method R-C\&W, and then the ones regarding the gradient projection solutions with $\ell_\infty$ constraint.

In Table \ref{tab:cwtotal}, we compare the $ASR$ and $ASR'$ performance of the vanilla C\&W, ComModel \cite{wang2020towards}, JPEGR \cite{shin2017jpeg} and our proposed R-C\&W. For the fairness of the comparison, we enforce that the compared methods have similar average distortions in $\ell_2$ sense. For vanilla C\&W, the $ASR$ is consistently 100\%, meaning that it is very powerful if no OSN transmission happens. However, the $ASR'$ quickly drops (to $10\sim30\%$) when the adversarial images are transmitted through Facebook, WeChat or QQ platforms. This maybe because C\&W produces AEs tightly close to the decision boundary of the target classifier, and hence, small disturbances caused by OSN transmission could easily pull them back to the original class. In contrast, R-C\&W still maintains 100\% $ASR$; but with much improved $ASR'$ results. For instance, in the case of Facebook transmission, our R-C\&W gains over 60$\%$ in $ASR'$ than vanilla C\&W when attacking the Inception-V3 model, which is also significantly better than the $ASR'$ results provided by the C\&W w/ ComModel and C\&W w/ JPEGR. This demonstrates that R-C\&W can significantly improve the robustness of AEs against the OSN transmission. ComModel and JPEGR focus more on improving the $ASR'$ performance; but hurt the $ASR$ performance severely. It should also be noted that, though ComModel and JPEGR achieve better $ASR'$ performance compared with vanilla C\&W, they are much inferior to our proposed R-C\&W. This is primarily due to our SIO model, capable of simulating the OSN manipulations more precisely, and the optimization formulation presented in Section \ref{section:cw} considering $ASR$ and $ASR'$ simultaneously.

We then evaluate the performance of the gradient projection approaches under the same $\ell_\infty$ constraint. The comparative results are presented in Tables \ref{tab:fgsm}-\ref{tab:mifgsm}. For the AEs transmitted over Facebook, WeChat or QQ, our proposed methods still achieve the best $ASR'$ performance almost for all the cases. For example, Our R-PGD achieves a significant improvement from 45$\%$ to 93$\%$ in $ASR'$ when attacking the VGG-19bn model over Facebook. Note that our robust methods outperform ComModel and JPEGR in terms of both $ASR$ and $ASR'$. Concerning $ASR$, our methods keep close to or even exceed vanilla methods (3.5\% gain with R-FGSM against ResNet-18), meanwhile maintaining the superiority of $ASR'$. Obviously, ComModel and JPEGR, which focus on improving $ASR'$, still suffer from a serious decline in $ASR$, especially for JPEGR, e.g., a 48$\%$ drop with vanilla PGD over WeChat when attacking VGG-19bn.

\begin{table*}[t!]
\setlength\tabcolsep{3pt}
  \centering
  \setcounter{table}{2}
  \caption{Robustness performance ($\%$) of AEs generated by R-FGSM (ours), FGSM integrated with ComModel\cite{wang2020towards}, JPEGR\cite{shin2017jpeg} and FGSM under the same $\ell_\infty$ constraint ($\epsilon=1/2/3/5$ for \texttt{ImageNet} over Facebook/\texttt{ImageNet} over QQ/\texttt{ImageNet} over WeChat/\texttt{Cifar-10} over Facebook respectively).}
  \scalebox{0.7}{
    \begin{tabular}{c|cc|cc|cc|cc|cc|cc}
    \hline
    \hline
  \multirow{4}{*}{Methods}     & \multicolumn{8}{c|}{Facebook}&\multicolumn{2}{c|}{WeChat}&\multicolumn{2}{c}{QQ}\Tstrut\\
   \cline{2-13}     & \multicolumn{6}{c|}{ImageNet}&\multicolumn{2}{c|}{Cifar-10}& \multicolumn{2}{c|}{ImageNet}&\multicolumn{2}{c}{ImageNet}\Tstrut\\
  \cline{2-13}     & \multicolumn{2}{c|}{VGG-19bn}  & \multicolumn{2}{c|}{ResNet-50}   & \multicolumn{2}{c|}{Inception-V3} &\multicolumn{2}{c|}{ResNet-18}& \multicolumn{2}{c|}{VGG-19bn} & \multicolumn{2}{c}{VGG-19bn} \Tstrut\\
\cline{2-13}          &  ASR$\uparrow$  & ASR$^\prime\uparrow$  &  ASR$\uparrow$  & ASR$^\prime\uparrow$   & ASR$\uparrow$  & ASR$^\prime\uparrow$  & ASR$\uparrow$  & ASR$^\prime\uparrow$ & ASR$\uparrow$  & ASR$^\prime\uparrow$& ASR$\uparrow$  & ASR$^\prime\uparrow$  \Tstrut  \\
    \hline
     R-FGSM (Ours) & \textbf{98.5} & \textbf{90.5}& 86.5 & \textbf{76.0}  & 79.5 &\textbf{77.0} & \textbf{57.5} & \textbf{53.5} & 89.0 &\textbf{71.0}& 98.0&\textbf{64.0}
   \Tstrut \\
     \hline
         FGSM w/ ComModel \cite{wang2020towards}& 95.0 &85.0 & 78.5 & 71.0  & 76.0 & 74.0 & 49.5 & 52.0 & 62.0&67.0& 96.0&56.0
         \Tstrut \\

   \hline
   FGSM w/ JPEGR\cite{shin2017jpeg}&79.5 &85.0&58.5 &71.5  &67.0 &73.0 &46.5 &44.5 & 55.0&68.0& 60.0&50.0
     \Tstrut \\
   \hline
    Vanilla FGSM   &  \textbf{98.5}&81.5&  \textbf{87.5}&66.5&  \textbf{81.5}&68.5& 54.0&48.0 & \textbf{99.0}&34.0& \textbf{100.0}&46.0
    \Tstrut  \\
    \hline
    \hline
    \end{tabular}}%
  \label{tab:fgsm}%
\end{table*}%

\begin{table*}[t!]
\setlength\tabcolsep{3pt}
  \centering
  \setcounter{table}{3}
  \caption{Robustness performance ($\%$) of AEs generated by R-PGD (ours), PGD integrated with ComModel\cite{wang2020towards}, JPEGR\cite{shin2017jpeg} and PGD under the same $\ell_\infty$ constraint ($\epsilon=1/2/3/5$ for \texttt{ImageNet} over Facebook/\texttt{ImageNet} over QQ/\texttt{ImageNet} over WeChat/\texttt{Cifar-10} over Facebook respectively).}
  \scalebox{0.7}{
    \begin{tabular}{c|cc|cc|cc|cc|cc|cc}
    \hline
    \hline
 \multirow{4}{*}{Methods}       & \multicolumn{8}{c|}{Facebook}&\multicolumn{2}{c|}{WeChat}&\multicolumn{2}{c}{QQ}\Tstrut\\
   \cline{2-13}     & \multicolumn{6}{c|}{ImageNet}&\multicolumn{2}{c|}{Cifar-10}& \multicolumn{2}{c|}{ImageNet}& \multicolumn{2}{c}{ImageNet}\Tstrut\\
  \cline{2-13}      & \multicolumn{2}{c|}{VGG-19bn}  & \multicolumn{2}{c|}{ResNet-50}   & \multicolumn{2}{c|}{Inception-V3} &\multicolumn{2}{c|}{ResNet-18} & \multicolumn{2}{c|}{VGG-19bn}  & \multicolumn{2}{c}{VGG-19bn} \Tstrut\\
\cline{2-13}          &  ASR$\uparrow$  & ASR$^\prime\uparrow$  &  ASR$\uparrow$  & ASR$^\prime\uparrow$   & ASR$\uparrow$  & ASR$^\prime\uparrow$   & ASR$\uparrow$  & ASR$^\prime\uparrow$ & ASR$\uparrow$  & ASR$^\prime\uparrow$& ASR$\uparrow$  & ASR$^\prime\uparrow$ \Tstrut  \\
    \hline
     R-PGD (Ours) & \textbf{99.0}&\textbf{93.0} &86.5&\textbf{88.5} &80.0&\textbf{81.5} & \textbf{100}&\textbf{99.0} & 92.0&\textbf{64.0}& \textbf{100}&\textbf{60.0}
   \Tstrut \\
   \hline

    PGD w/ ComModel\cite{wang2020towards}  &85.5&64.5&78.0&60.0 &70.0&61.0 &96.5&88.0  &56.0&39.0 &96.0&26.0 \Tstrut\\
    \hline
            PGD w/ JPEGR\cite{shin2017jpeg}&68.5&89.5&58.0&81.5&56.5&79.5&75.5&50.0 &52.0&58.0&54.0&26.0 \Tstrut\\
    \hline
    Vanilla PGD   & 98.0&45.0 & \textbf{94.0}&43.5&  \textbf{85.5}&53.0 & \textbf{100}&77.5  &\textbf{100}&20.0&\textbf{100}&8.0
    \Tstrut  \\
    \hline
    \hline
    \end{tabular}}%
  \label{tab:pgd}%
\end{table*}%

\begin{table*}[t!]
\setlength\tabcolsep{3pt}
  \centering
  \setcounter{table}{4}
  \caption{Robustness performance ($\%$) of AEs generated by R-MIFGSM (ours), MIFGSM integrated with ComModel\cite{wang2020towards}, JPEGR\cite{shin2017jpeg} and MIFGSM under the same $\ell_\infty$ constraint ($\epsilon=1/2/3/5$ for \texttt{ImageNet} over Facebook/\texttt{ImageNet} over QQ/\texttt{ImageNet} over WeChat/\texttt{Cifar-10} over Facebook respectively).}
  \scalebox{0.7}{
    \begin{tabular}{c|cc|cc|cc|cc|cc|cc}
    \hline
    \hline
  \multirow{4}{*}{Methods}   & \multicolumn{8}{c|}{Facebook}&\multicolumn{2}{c|}{WeChat}&\multicolumn{2}{c}{QQ}\Tstrut\\
    \cline{2-13}    & \multicolumn{6}{c|}{ImageNet}&\multicolumn{2}{c|}{Cifar-10}& \multicolumn{2}{c|}{ImageNet}& \multicolumn{2}{c}{ImageNet}\Tstrut\\
  \cline{2-13}    & \multicolumn{2}{c|}{VGG-19bn}  & \multicolumn{2}{c|}{ResNet-50}   & \multicolumn{2}{c|}{Inception-V3} &\multicolumn{2}{c|}{ResNet-18} & \multicolumn{2}{c|}{VGG-19bn}  & \multicolumn{2}{c}{VGG-19bn} \Tstrut\\
   \cline{2-13}      &  ASR$\uparrow$  & ASR$^\prime\uparrow$  &  ASR$\uparrow$  & ASR$^\prime\uparrow$    & ASR$\uparrow$  & ASR$^\prime\uparrow$ & ASR$\uparrow$  & ASR$^\prime\uparrow$ & ASR$\uparrow$  & ASR$^\prime\uparrow$ & ASR$\uparrow$  & ASR$^\prime\uparrow$  \Tstrut  \\
    \hline
     R-MIFGSM (Ours)  &98.5&\textbf{94.5}&93.0&\textbf{88.0} &92.5&\textbf{84.5} &88.0&\textbf{82.0}  &98.0&80.0&\textbf{100}&\textbf{70.0} \Tstrut\\
    \hline

    MIFGSM w/ ComModel\cite{wang2020towards}  &98.0&84.5&91.5&77.0 &89.5&81.0 &83.5&79.0 &79.0&65.0&\textbf{100}&62.0 \Tstrut\\
    \hline
      MIFGSM w/ JPEGR\cite{shin2017jpeg}&92.0&93.0&81.0&85.5&81.5&83.5&75.5&64.0&79.0&\textbf{86.0}&92.0&62.0 \Tstrut\\
    \hline
    Vanilla MIFGSM   &  \textbf{99.5}&77.5& \textbf{98.5}&67.0 &  \textbf{97.5}&75.5 & \textbf{90.0}&79.0 & \textbf{100}&34.0& \textbf{100}&24.0
    \Tstrut  \\
    \hline
    \hline
    \end{tabular}
    }%
  \label{tab:mifgsm}%
\end{table*}%

\begin{center}
\begin{figure*}[t!]
\centering{\includegraphics[width=0.7\textwidth]{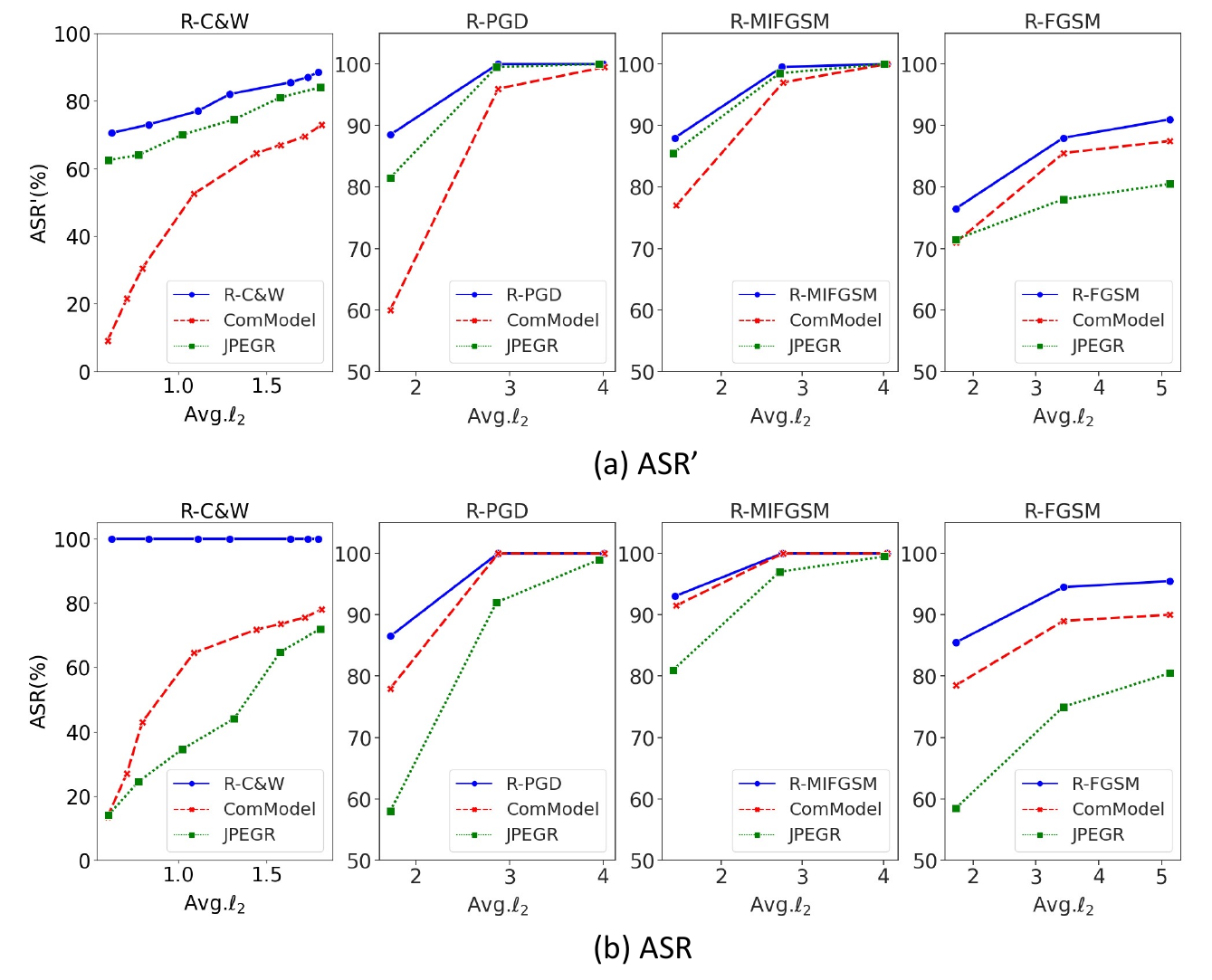}}
\caption{The $ASR'$ (a) and $ASR$ (b) comparisons with respect to different Avg.$\ell_2$ perturbation against the pre-trained ResNet-50 classification model.}
\label{fig:asrl2}
\end{figure*}
\end{center}

More performance comparisons of $ASR'$ and $ASR$ with respect to different magnitudes of adversarial perturbations are shown in Fig.\ref{fig:asrl2}. Our robust attack methods achieve superior $ASR'/ASR$ results especially in small perturbations. For example, when $Avg.\ell_2<1$ for R-C\&W and $Avg.\ell_2<2$ for R-FGSM/R-PGD/R-MIFGSM, the performance gains over ComModel and JPEGR are quite significant. With larger distortion budgets, AEs can be far away from the decision boundary, which helps AEs still locate outside the decision boundary even after OSN transmissions. Thus the robustness of AEs produced by ComModel and JPEGR could be improved for larger distortion budgets, narrowing the $ASR'$ gains of our methods. Thanks to our SIO model and the well-designed optimization loss functions in (\ref{eq:cw3}) and (\ref{eq:jointce2}), our proposed methods outperform ComModel and JPEGR in $ASR$ and $ASR'$ simultaneously almost for all the cases.

Furthermore, it would also be interesting to find out what kind of images benefit more from our approach. To this end, we conduct an error analysis on both successful and unsuccessful AEs after OSN transmissions. As can be concluded from Appendix \ref{sec:apa3} and \ref{section:appcl}, our method is more effective in improving the robustness of those AEs that suffer from smaller OSN noises. Additionally, we observe that after being transmitted over OSN, the successful AEs generated by our method have a relatively high confidence level.

\subsection{Simulation performance of SIO}\label{section:simusio}
In addition to the comparison regarding the robustness, we present more results on the SIO model, which is one of the key modules in our proposed scheme.

We first examine the effects of the residual learning, the differentiable JPEG layer and the SCSE blocks in the SIO network. Four configured models are compared: SCSE U-Net, SCSE U-Net with the residual learning denoted by ``SCSE U-Net+Res.'', SCSE U-Net+Res. appended with the differential JPEG layer shortened as ``SCSE U-Net+Res.+JPEG'' (a.k.a our SIO network) and SIO without SCSE blocks (SIO w/o SCSE). These models are separately re-trained on the same dataset. The simulation performance comparison among these models is tabulated in Table \ref{tab:modelscompare}, where MSE loss, average PSNR and SSIM between the simulated noise $\delta_g=\mathcal{G}_\theta(\mathbf{x})-\mathbf{x}$ and ground truth noise $\delta=\mathrm{OSN}(\mathbf{x})-\mathbf{x}$ of 200 images in the validation set of \texttt{ImageNet} are used as metrics. It can be seen that the average PSNR gain by the residual learning reaches 3.13 dB, demonstrating its effectiveness in enhancing the simulation capability of the OSN noise. Also, the incorporation of the differentiable JPEG layer can further boost the simulation performance significantly. Once we remove SCSE blocks from SIO, the simulation performance in terms of PSNR drops by 0.3 dB, demonstrating that SCSE blocks also improve the simulation attainment. As for the attack ability, better OSN simulation performance generally leads to more robust AEs, as can be observed from Table \ref{tab:modelscompare} and \ref{tab:simatt}.

\begin{figure*}[t!]
\centering\includegraphics[width=0.9\textwidth]{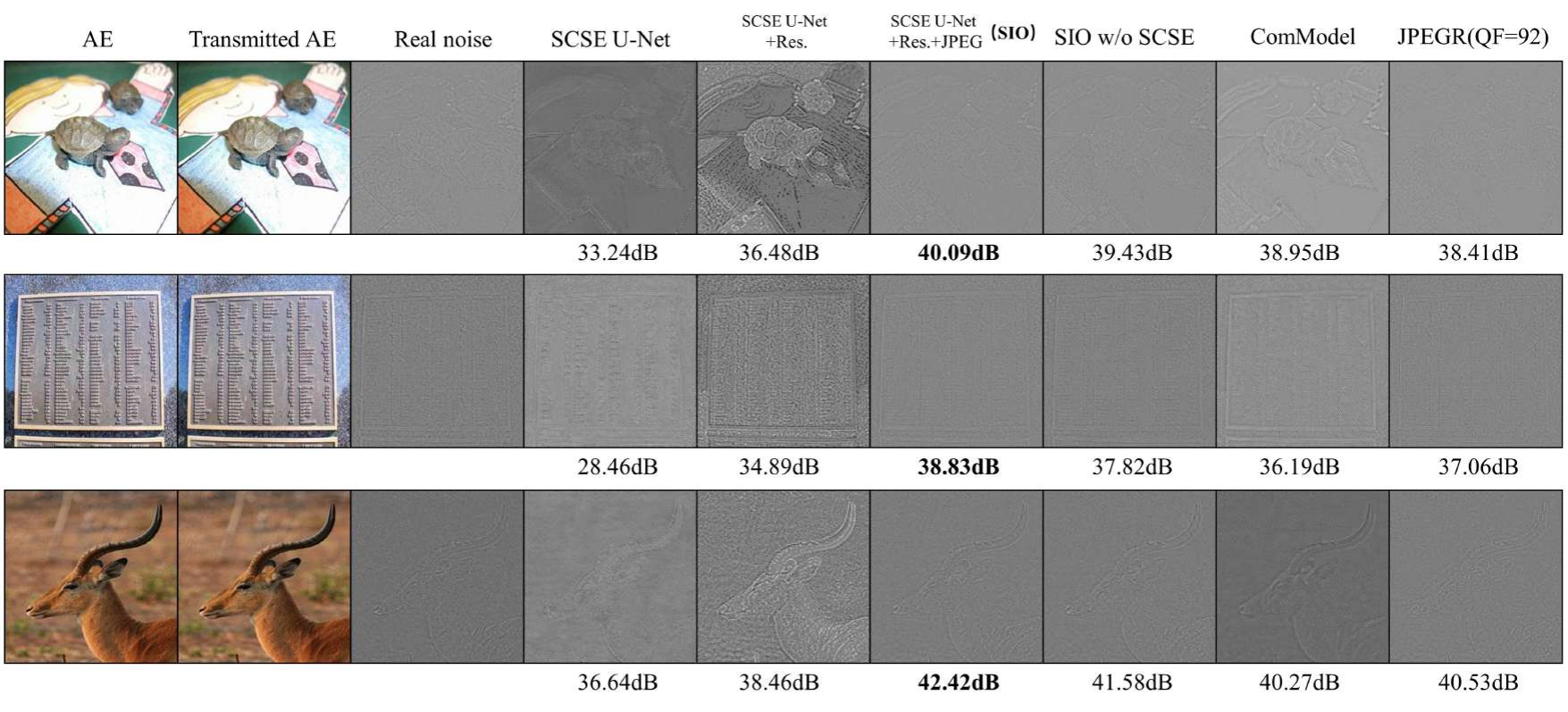}
\caption{Visualization of the simulated noise of Facebook by different network architectures. The 1\textsuperscript{st} column lists AEs generated by vanilla attack methods and randomly sampled from \texttt{ImageNet} validation set. Transmitted AEs over Facebook are shown in the 2\textsuperscript{nd} column. The 3\textsuperscript{rd} column shows the normalized real noises, which are obtained by subtracting the second column from the first one. The remaining columns visualize the noises produced by the corresponding network architectures.}
\label{fig:ablation}
\end{figure*}

In Fig.\ref{fig:ablation}, we also show the visual comparison among the noises generated by the above three models. It can be observed that the residual learning does help the network better learn the structural noise caused by Facebook. However, it should also be noticed that the noises produced by ``+ Res.'' are too sharp and still far away from the real ones, partially because it is difficult for a DNN to simulate the unique pattern caused by the JPEG compression. Augmented with the differentiable JPEG layer, the resulting noises are visually much more similar to the real ones.

\begin{figure}[ht]
\centering\includegraphics[width=0.7\textwidth]{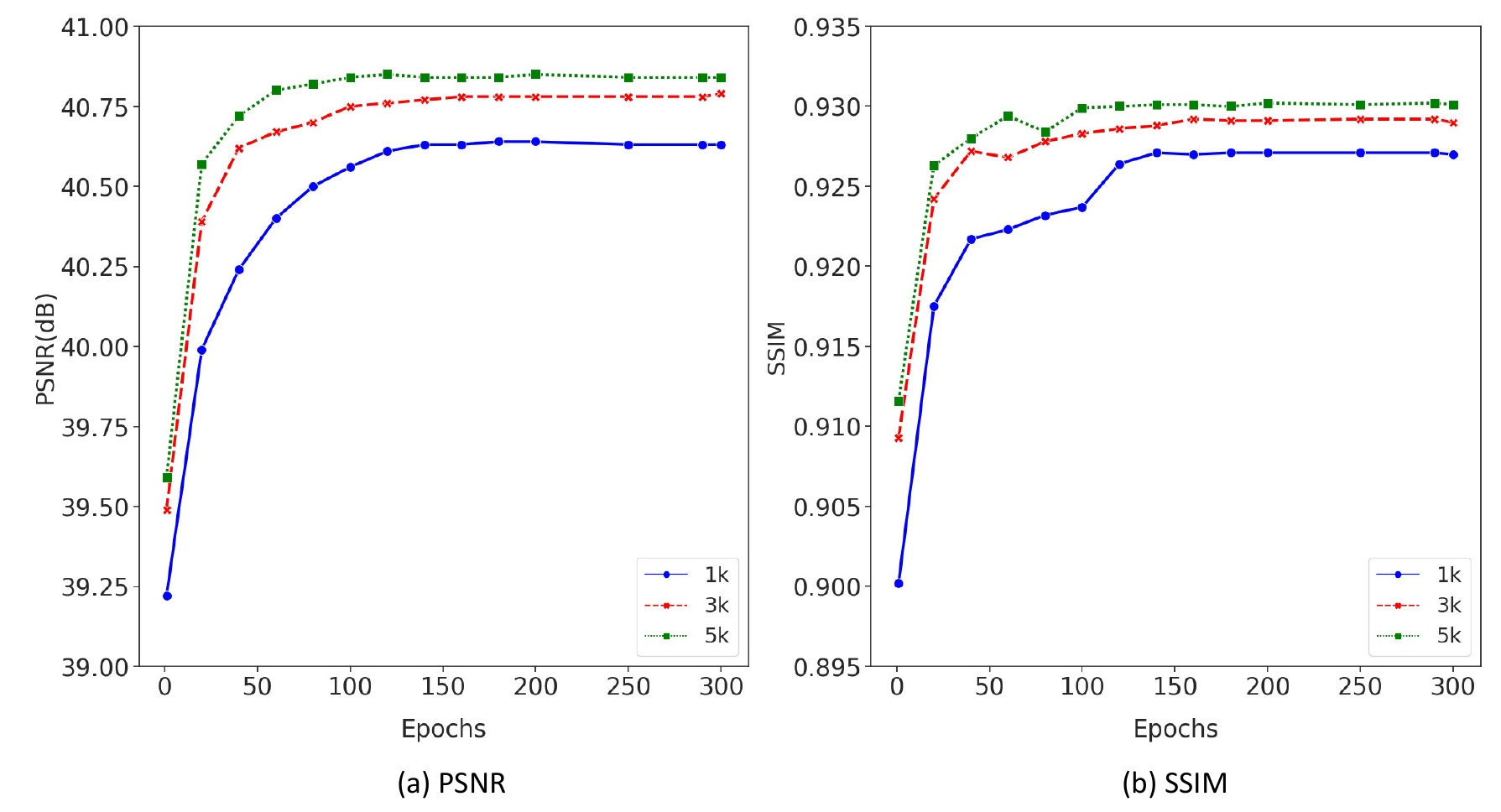}
\caption{Simulation performance of SIOs trained with 1000, 3000 and 5000 pairs of images, respectively.}
\label{fig:losscurve}
\end{figure}

 Furthermore, we study the influence of the number of training data on the simulation performance of the resulting SIO. In Fig.\ref{fig:losscurve}, we show the simulation performance in terms of PSNR and SSIM with 1000, 3000, and 5000 pairs of images, respectively, where the other training settings are kept the same. Here, the test set contains 200 AEs based on images randomly sampled from the validation set of \texttt{ImageNet}. It can be observed that 3000 pairs of training images lead to big improvements, compared with the case of 1000 pairs. However, further increasing the number of pairs to 5000 only results in almost indistinguishable improvements, but with heavier training costs. Therefore, it is sufficient to adopt 3000 pairs of training images, achieving a good trade-off between the simulation performance and the training cost.

Moreover, our SIO model not only exhibits superior simulation performance on Facebook compared to ComModel and JPEGR, but also surpasses them on WeChat and QQ. For further experimental details, please refer to Appendix \ref{sec:a1}.

\begin{table}[t!]
  \centering
  \caption{The performance of simulated noise over Facebook by different network architectures. $^+$ Higher is better. $^-$ lower is better.}
  \scalebox{0.65}{
    \begin{tabular}{c c c c }
      \toprule
    \toprule
    Networks &  PSNR$^+$&  SSIM$^+$&  MSE$^-$ \\
    \midrule
    SCSE U-Net& 34.22 & 0.5893 & 28.33  \\
    \midrule
    SCSE U-Net+Res. & 37.35 & 0.8298 & 14.29 \\
    \midrule
     SCSE U-Net+Res.+JPEG (SIO) & \textbf{40.63} & \textbf{0.9270} & \textbf{6.85} \\
    \midrule
    SIO w/o SCSE & 40.33 & 0.9193 & 7.26 \\
    \midrule
    ComModel\cite{wang2020towards} & 39.42 & 0.8989 & 8.31 \\
 \midrule
    JPEGR\cite{shin2017jpeg} & 39.05 & 0.9051 & 9.94 \\
    \bottomrule
       \bottomrule
    \end{tabular}%
    }
  \label{tab:modelscompare}%
\end{table}%

\begin{table*}[t!]
  \centering
  \caption{The robustness performance ($\%$) comparison of AEs transmitted over Facebook against ResNet-50, where AEs are generated by different solutions or network architectures. For each module, a line search is performed to find the constant $\lambda\in[0.1,0.9]$ that yields AEs of the \textbf{highest average $ASR^\prime$} with a step size of 0.1. Thus the $ASR$ achieved by SIO is not the best among other architectures. Note the $l_\infty$ constraint of FGSM/PGD/MIFGSM is $\epsilon=1$, and module `-' represents generating AEs by vanilla FGSM/PGD/MIFGSM/C\&W method.}
  \scalebox{0.7}{
    \begin{tabular}{c|c|c|c|c|c|c|c|c|c}
      \hline
    \hline
    Solutions & \multicolumn{2}{c|}{FGSM} & \multicolumn{2}{c|}{PGD} & \multicolumn{2}{c|}{MIFGSM} & \multicolumn{3}{c}{Lagrange-form}\Tstrut \\
    \hline
    Modules & ASR$\uparrow$   & ASR$^\prime\uparrow$  & ASR$\uparrow$   & ASR$^\prime\uparrow$  & ASR$\uparrow$   & ASR$^\prime\uparrow$  & ASR$\uparrow$   & ASR$^\prime\uparrow$  & Avg.$l_2$ \Tstrut\\
    \hline
    - & 87.5  & 66.5  & 94.0  & 43.5  & 98.5  & 67.0  & 100   & 19.0  &0.784 \Tstrut\\
    \hline
    SCSE U-Net & 87.0  & 72.0  & 89.0  & 56.0  & 94.5  & 74.0  & 100   &36.5  &0.777 \Tstrut\\
    \hline
    SCSE U\_Net+Res. & 89.0  & 73.5  & 91.5  & 60.5  & 98.5  & 77.5  & 100   &45.0   &0.777 \Tstrut\\
    \hline
SCSE U\_Net+Res.+JPEG \textbf{(SIO)} & 86.5  & \textbf{76.0} & 86.5  & \textbf{88.5} & 93.0  & \textbf{88.0} & 100   & \textbf{70.5} &0.747 \Tstrut\\
    \hline
    SIO w/o SCSE & 85.0  & 74.0  & 94.0  & 84.5  & 95.0  & 86.5  & 100 &65.0 &0.772 \Tstrut\\
    \hline
    \hline
    \end{tabular}%
}
  \label{tab:simatt}%
\end{table*}%

\subsection{The impact of optimization parameters on performance}\label{section:ablation}

\begin{center}
\begin{figure*}[t!]
\centering{\includegraphics[width=0.7\textwidth]{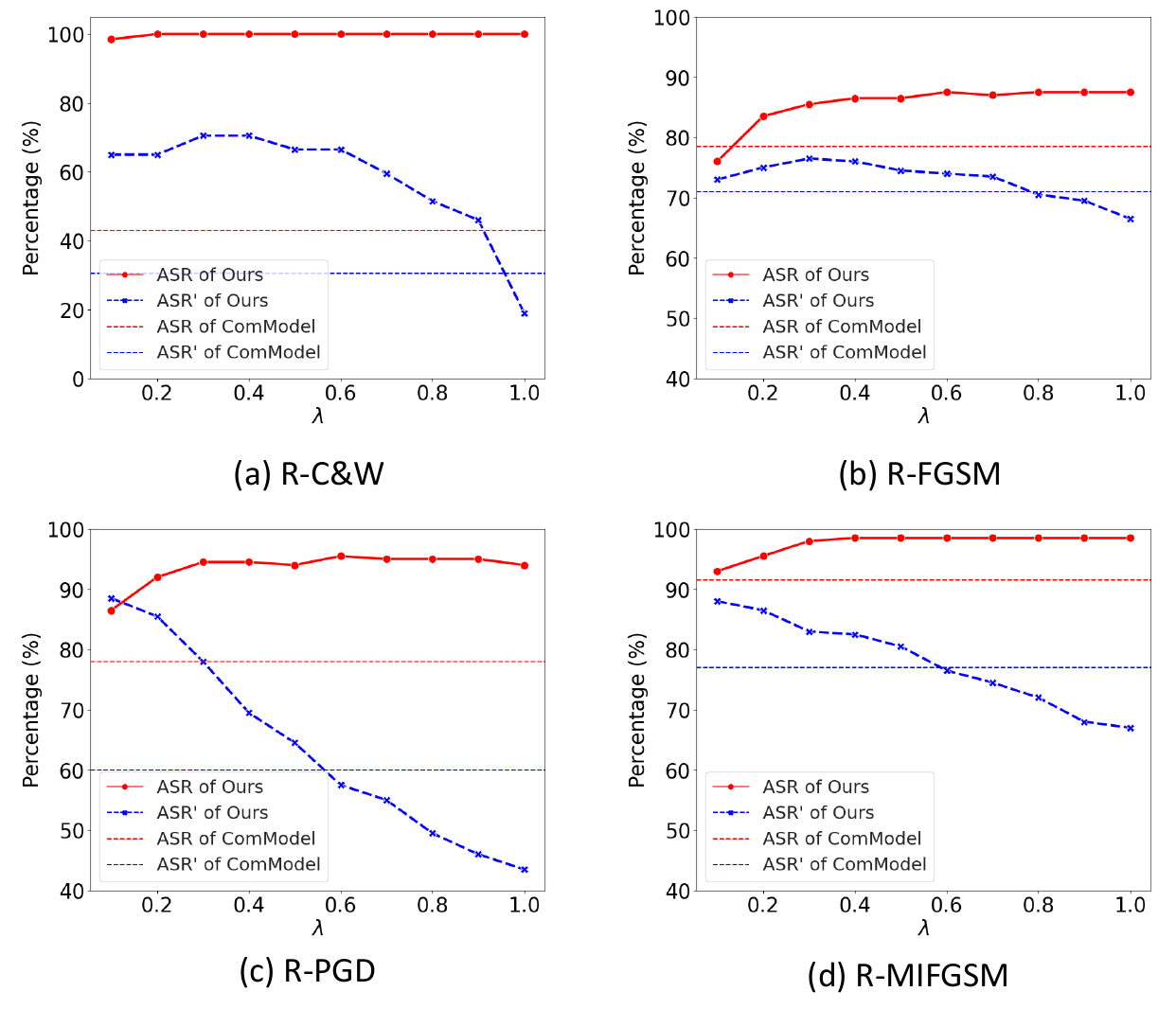}}
\caption{The influence of $\lambda$ on the robustness performance against transmission over Facebook when attacking ResNet-50.}
\label{fig:lambda}
\end{figure*}
\end{center}

We also study the influence of $\lambda$ in (\ref{eq:cw3}) and (\ref{eq:jointce2}) on the $ASR$ and $ASR^\prime$ performance. Specifically, we compare the performance of R-C\&W/R-FGSM/R-PGD/R-MIFGSM over Facebook when changing the value of $\lambda$ within the range of $(0,1]$, as shown in Fig.\ref{fig:lambda}. It should be noted that R-C\&W/R-FGSM/R-PGD/R-MIFGSM degenerate into vanilla attack methods (C\&W/FGSM/PGD/MIFGSM respectively) when $\lambda=1$. For comparison purposes, we use horizontal straight lines to indicate $ASR/ASR'$ of ComModel and JPEGR. Since these methods do not involve the parameter $\lambda$, their $ASR/ASR'$ with respect to $\lambda$ are constants when other attack settings are intact. For our methods, the increase of $\lambda$ would lead to the improvement of $ASR$ and the decrease of $ASR^\prime$ in general. Particularly, the proposed methods achieve higher $ASR^\prime$ consistently than corresponding vanilla methods ($\lambda=1$). In addition, when compared with ComModel, for a wide range of $\lambda$, $ASR$ and $ASR^\prime$ obtained by our methods are higher. While for JPEGR, our methods outperform it in terms of $ASR'$ when $\lambda$ is relatively small and always reach substantially higher $ASR$ within the range of (0,1]. Based upon these observations, it is suggested to set a relatively small $\lambda$ to achieve satisfactory $ASR$ and $ASR'$ performance at the same time.

\begin{table*}[t!]
  \centering
  \caption{The robustness performance (\%) of AEs generated by comparative methods for Facebook, while transmitted over Facebook, WeChat, and QQ, respectively. Here, VGG19bn is the target model. The superscript $*$ means that the simulated OSN and the transmitted one are the same.}
  \scalebox{0.7}{
    \begin{tabular}{c|c|c|c|c|c|c}
    \hline\hline
     &   &  & & \multicolumn{3}{c}{Transmitted OSNs} \\
\cline{5-7}Methods          & Simulated OSNs      & Avg.$l_2/l_\infty$ &  ASR   & Facebook & WeChat & QQ  \\
\cline{5-7}          &       &   &    & ASR$^\prime\uparrow$  & ASR$^\prime\uparrow$  & ASR$^\prime\uparrow$ \\
    \hline
    Vanilla CW & -&0.591 & 100   & 12.0    & 16.5  & 12.0 \\
CW w/ JPEGR & Facebook & 0.597 &  14.5  & 65.0 &19.2  &15.5 \\
      CW w/ ComModel & Facebook & 0.602 &  31.0  &21.5 & 16.4 &14.5 \\
            \textbf{R-C\&W (Ours)}& Facebook & 0.601 & 100   & \textbf{70.0}$^*$ & \textbf{37.5} & \textbf{46.5} \\
    \hline
    
    Vanilla FGSM & -&1& 98.5  & 81.5  & 31.0    & 36.0 \\
         FGSM w/ JPEGR & Facebook & 1 & 79.5   & 85.0 & 30.6 & 42.0\\
       FGSM w/ ComModel & Facebook &1  & 95.0 &  85.0& 32.3 & 45.0\\
  \textbf{R-FGSM (Ours)} &Facebook &    1   & 98.5  & \textbf{90.5}$^*$  & \textbf{43.0} & \textbf{46.5} \\
    \hline
   
    Vanilla PGD & -&{1} & 98.0    & 45.0    & 19.5  & 21.0 \\
             PGD w/ JPEGR & Facebook & 1 &  68.5 & 89.5  & 20.4 & 24.5\\
         PGD w/ ComModel & Facebook & 1 & 85.5  & 64.5&  21.2 & 25.0\\
\textbf{R-PGD (Ours)} &Facebook &    1   & 99.0    & \textbf{93.0}$^*$ & \textbf{25.0} & \textbf{26.0} \\
    \hline
   
    Vanilla MIFGSM & -&{1}& 99.5  & 77.5  & 26.0    & 26.5 \\
                 MIFGSM w/ JPEGR & Facebook & 1& 92.0    &93.0 & 23.5  & 34.5\\
         MIFGSM w/ ComModel & Facebook & 1 &98.0 &  84.5  & 28.6 & 35.0 \\
  \textbf{R-MIFGSM (Ours)} & Facebook &    1   & 98.5  & \textbf{94.5}$^*$ & \textbf{36.5} & \textbf{41.0} \\
    
    \hline
    \hline
    \end{tabular}%
    }
  \label{tab:generalibility}%
\end{table*}%

\subsection{Performance evaluation on cross-platform scenarios}

In the above experiments, it is assumed that a specific OSN platform is known in advance. In reality, the transmission platform could be accurately predicted using the historical propagation trajectory, or it can be explicitly specified, e.g., users can only use Dingding or WeChat for the file transfer, mainly due to security concerns.

We now consider more complex cases, where the OSN used for training the SIO and the actual transmission OSN are different, namely, the cross-platform scenarios. We report the attack performance of AEs generated by comparative methods for Facebook, and then transmitted over Facebook, WeChat or QQ, respectively. Specifically, in Table \ref{tab:generalibility}, we give the ASR and ASR$^\prime$ performance under different attack settings. It can be seen that, though the enhancement of ASR$^\prime$ obtained by our methods drops for WeChat and QQ platforms, it is still significantly better than those of ComModel and JPEGR; e.g. up to 34.5\% gain by R-C\&W over QQ, but only 2.5\%/3.5\% improvements by other two techniques. This implies that our proposed method trained with one specific OSN platform can still be generalized to other platforms. Certainly, how to achieve better generalization, especially when the gap between the training and testing platforms is large, would be a very interesting problem to be explored in the future.

\begin{table*}[t]
  \centering
  \caption{The ASR$^\prime$(\%) of vanilla and our proposed methods against ResNet-50 with or without advanced defense approaches. Here the OSN platform is selected to be Facebook.}
  \scalebox{0.7}{
    \begin{tabular}{c|c|c|c|c|c|c|c|c|c|c}
    \hline\hline
    
    \multirow{2}{*}{Methods} & \multirow{2}{*}{Avg.$l_2/l_\infty$} & \multirow{2}{*}{ASR} & \multicolumn{8}{c}{ASR$^\prime\uparrow$} \Tstrut\\
\cline{4-11}          &       &       & w/o defense & Bit-Red\cite{featureSqueeze} & R\&P\cite{xiemitigating} & JPEG(75) & FD\cite{liu2019feature}    & ComDefend\cite{jia2019comdefend}
 & NRP\cite{naseer2020self} &Average \Tstrut\\

    \hline
    Vanilla CW & 0.784 & 100   & 19.0        & 8.0     & 10.5  & 10.0    & 34.5  & 12.5  & 24.0 & 16.9\Tstrut\\
    
    R-CW  & 0.747 & 100   & \textbf{70.5}     & \textbf{19.5} & \textbf{35.5} & \textbf{21.5} & \textbf{39.0} & \textbf{23.0} & \textbf{28.5} &\textbf{33.9}\Tstrut\\
    \hline
    
    Vanilla FGSM & {1} & 87.5  & 66.5      & 36.5  & 42.0    & 42.5  & 40.5  & 32.5  & 30.0 &41.5\Tstrut\\
   R-FGSM &     1  & 86.5  & \textbf{76.0}     & \textbf{42.5} & \textbf{43.5} & \textbf{49.0} & \textbf{45.0} & \textbf{39.5} & \textbf{33.0} & \textbf{46.9}\Tstrut\\\hline 
   Vanilla PGD &  {1}     & 94.0    & 43.5          & 25.0& 23.5  & 19.5  & 41.0 & 21.0    & 31.0& 29.2\Tstrut\\
    R-PGD &   1    & 86.5  & \textbf{88.5}     & \textbf{41.5} & \textbf{41.5} & \textbf{39.5} & \textbf{43.5}    & \textbf{36.0} & \textbf{34.0} &\textbf{46.4}\Tstrut\\\hline
Vanilla MIFGSM & {1}      & 98.5  & 67.0        & 32.0    & 37.0    & 34.5  & 42.0    & 28.0    & 30.5 &38.7\Tstrut\\
   R-MIFGSM &    1   & 93.0    & \textbf{88.0}     & \textbf{44.5} & \textbf{52.0} & \textbf{50.0} & \textbf{44.5} & \textbf{39.0} & \textbf{34.0} & \textbf{50.3}\Tstrut\\
    \hline
    \hline
    \end{tabular}%
    }
  \label{tab:defense}%
\end{table*}%
\subsection{Performance evaluation against defense methods}
We also evaluate the effectiveness of the proposed attack methods under some pre-processing based defense approaches 
\cite{featureSqueeze,xiemitigating,liu2019feature,jia2019comdefend,naseer2020self}. Specifically, these defense approaches pre-process the transmitted AEs before feeding them to the classifier by using bit-depth reduction (Bit-Red)\cite{featureSqueeze}, random resizing and padding (R\&P)\cite{xiemitigating}, JPEG compression with $QF=75$, feature denoising (FD)\cite{liu2019feature}, a DNN-based compression model (ComDefend)\cite{jia2019comdefend} or a neural representation purifier (NRP)\cite{naseer2020self}. As shown in Table \ref{tab:defense}, we test the ASR$^\prime$ of AEs with or without defense over Facebook. These AEs are generated on ResNet-50 with the SIO for Facebook. It can be observed that in all cases, our methods are more robust against defenses than vanilla attack methods. For example, our R-C\&W obtains 25\% gains in ASR$^\prime$ than vanilla C\&W method when facing the R\&P defense.           

It should also be noted that this work focuses on enhancing the robustness of AEs against OSN transmissions (but not particularly on defeating the defense approaches), and our robust design can work well with many existing attack methods. Once these attack methods are improved against defenses, the robust versions will naturally inherit such capabilities.

\section{Conclusion}\label{section:con}
In this paper, we have designed a new framework for generating robust AEs that simultaneously maintain the attack capabilities before and after the OSN transmissions. Our proposed framework consists of a new differentiable network for simulating the OSN and a novel optimization formulation with constraints specifically addressing these attack capabilities. Extensive experiments conducted over Facebook, WeChat and QQ have demonstrated that our attack methods produce more robust AEs than existing approaches, especially under small distortion constraints. Furthermore, we have built a public dataset containing more than 10,000 pairs of AEs processed by Facebook, WeChat or QQ, facilitating future research on generating robust AEs.

\bibliographystyle{ACM-Reference-Format}
\bibliography{sample-base}

\appendix

\section{Appendix}

\subsection{The impact of the parameters used in the training stage}\label{sec:appa2}

The attack parameters used in the training stage for generating AEs do affect both the simulation and robustness performance; but their impacts are negligible. As a result, the parameters given in Table \ref{tab:attpara} shall be enough for the performance evaluations. We now give some specific examples. We train three SIO models using AEs generated with FGSM under $l_\infty$ constraints of different parameters, i.e. $\epsilon=1,2,3$, respectively. As depicted in Table \ref{tab:modelscompareAEs}, the simulation performance of the SIO model just slightly changes from 40.82 dB to 40.70 dB, when $\epsilon$ increases. We also have observed similar behaviors when changing the other parameters in a reasonable range. Furthermore, due to the relatively small variations in simulation performance, networks trained with different parameters have a minor impact on the ultimate attack effectiveness as well. For instance, the maximum fluctuation in ASR$^\prime$ is only 1\% when the SIO is incorporated with R-PGD (see the fourth column of Table \ref{tab:modelscompareAEs}).

\subsection{The error analysis on the successful and unsuccessful AEs after OSN transmission}\label{sec:apa3}

To figure out the characteristics of images that our methods are more effective for, we conduct error analysis on AEs generated by our methods. In general, our method works better on those images that suffer from smaller OSN noises. Apparently, the noise added by OSN transmission is image-dependent, and hence, different images could lead to different levels of OSN noise. To support this statement, we have conducted the following experiments. 

We construct a test set $\mathbb{A}$, which consists of 200 AEs generated by our R-FGSM method based on clean images randomly selected from the validation set of \texttt{ImageNet}. We then divide $\mathbb{A}$ into two subsets $\mathbb{A}_s$ and $\mathbb{A}_f$, according to whether or not the AEs can still fool the target model after being transmitted over Facebook. Also, we can readily obtain the actual OSN noise and simulation error for these two subsets, by calculating the average values of $\mathrm{MSE}(\mathbf{x}^{*}-\mathrm{OSN}(\mathbf{x}^{*}))$ and $\mathrm{MSE}(\mathcal{G}_\theta(\mathbf{x}^{*})-\mathrm{OSN}(\mathbf{x}^{*}))$, respectively, where $\mathbf{x}^{*}$ denotes an AE. As shown in Table \ref{tab:errorany}, in all cases, the OSN noise and simulation error associated with $\mathbb{A}_s$ are lower than those of $\mathbb{A}_f$. This is because the smaller the OSN noise in general leads to the smaller the simulation error, which in turn results in a higher attack success rate after the OSN transmission. Therefore, our method tends to be more effective in improving the robustness of those AEs whose associated OSN noises are relatively small.

\subsection{The confidence level comparison of successful and unsuccessful AEs after OSN transmission}\label{section:appcl}

An AE generated by our method that can still attack successfully after OSN transmission often does have a relatively high Confidence Level ($\mathrm{CL}$) to cheat the target model. To support this statement, we have conducted the following analysis.

Clearly, if the $\mathrm{CL}$ to cheat the target model is high, the successful AEs should be predicted with a high $\mathrm{CL}$ of $\hat{y}$ or a low $\mathrm{CL}$ of $y$, where $y$ represents the ground truth label of our generated image $\mathbf{x}^*$ and $\hat{y}=\mathop{\max}\nolimits_{j\neq y}\mathcal{Z}(\hat{\mathbf{x}})[j]$. Here, $\hat{\mathbf{x}}=\mathrm{OSN}(\mathbf{x}^*)$, $\mathcal{Z}(\cdot)=[z_0,z_1,...,z_{c-1}]$ stands for the logits output vector of the penultimate layer in the target model and $\mathcal{Z}(\hat{\mathbf{x}})[j]$ returns the $j$th element of $\mathcal{Z}(\hat{\mathbf{x}})$. For a comparison, we also calculate the $\mathrm{CL}$ of unsuccessful AEs that are also crafted by our method but are correctly classified after the OSN transmission.

In light of such a motivation, we formally define a metric Average Confidence Level ($\mathrm{ACL}$). Specifically, given an image set $\mathbb{O}$ with $|\mathbb{O}|=M$ and based on the classic definition of confidence \cite{ishida2018binary}, $\mathrm{ACL}$ is expressed as:

\begin{large}
   \begin{equation}\label{eq:conl}
\begin{gathered}
\mathrm{ACL}(v,\mathbb{O})=
\begin{cases}
\frac{1}{M}\sum\nolimits_{k=0}^{M-1}\frac{\exp(\mathcal{Z}(\hat{\mathbf{x}}_{k})[{\color{red}y_{k}}])}{\sum\nolimits^{c-1}_{j=0}\exp(\mathcal{Z}(\hat{\mathbf{x}}_{k})[j])}\ \text{if $v$ is } \mathrm{True},
\\
 \frac{1}{M}\sum\nolimits_{k=0}^{M-1}\frac{\exp(\mathcal{Z}(\hat{\mathbf{x}}_{k})[{\color{red}\hat{y}_{k}}])}{\sum\nolimits^{c-1}_{j=0}\exp(\mathcal{Z}(\hat{\mathbf{x}}_{k})[j])}\  \text{else,}
\end{cases}
  \hat{\mathbf{x}}_{k}\in\mathbb{O},
\end{gathered}
\end{equation}
\end{large} where $c$ is the number of classes. By setting the parameter $v$ to $\mathrm{True}$ or $\mathrm{False}$, we can obtain the average $\mathrm{CL}$ corresponding to $y$ or $\hat{y}$, respectively. 

Subsequently, we construct two subsets $\mathbb{O}_s$ and $\mathbb{O}_f$, which are formed by the Facebook transmitted AEs from $\mathbb{A}_s$ or $\mathbb{A}_f$, respectively. Note that $\mathbb{A}_s$ or $\mathbb{A}_f$ are defined above in Appendix \ref{sec:apa3}. By definition, AEs in $\mathbb{O}_s$ can deceive the target model, whereas the ones in $\mathbb{O}_f$ cannot. Then we calculate $\mathrm{ACL}$ of these two subsets using Eq.(\ref{eq:conl}) above, and the results are compiled into Table \ref{tab:c5}. It can be seen that successful AEs correspond to low $\mathrm{ACL}$ of being classified as the ground truth label (e.g. $\mathrm{ACL}(\mathrm{True},\mathbb{O}_s)<6.5\%$ for all models), which in turn implies that the $\mathrm{CL}$ values of other wrong classes are high. In other words, there is a very high probability to deceive the target model. Similarly,  the $\mathrm{ACL}$ of successful AEs to be predicted as the wrong class $\hat{y}$ is also relatively high (e.g. $\mathrm{ACL}(\mathrm{False},\mathbb{O}_s)=80.24\%$ for Inception-V3). Therefore, AEs generated by our method that still preserve attack capability after being transmitted over Facebook do have relatively high $\mathrm{CL}$s.

\begin{table}[t]
  \centering
  \caption{The performance of simulated noise over Facebook by SIO networks trained on AEs generated with different parameters, and the attack performance of our methods integrated with these SIO models. The attack target model is ResNet-50 pre-trained in the ImageNet dataset. The test set for simulation consists of AEs generated by FGSM under the constraint of $\epsilon=1$.}
   \scalebox{0.85}{
    \begin{tabular}{c|c|cc|cc|cc|cc}
    \hline\hline
    \multirow{2}{*}{Networks} & Simulation & \multicolumn{2}{c|}{R-FGSM($\epsilon$=1)} & \multicolumn{2}{c|}{R-PGD($\epsilon$=1)} & \multicolumn{2}{c|}{R-MIFGSM($\epsilon$=1)} & \multicolumn{2}{c}{R-C\&W($l_2\approx$0.747)} \\
\cline{2-10}          & PSNR$\uparrow$  & ASR   & ASR$^\prime\uparrow$   & ASR   & ASR$^\prime\uparrow$  & ASR   & ASR$^\prime\uparrow$  & ASR   & ASR$^\prime\uparrow$ \\
    \hline
    SIO w/ FGSM ($\epsilon$=1) &       \textbf{40.82}   &         85.5  &         \textbf{76.0}  &         91.0  &         \textbf{90.0}&         95.5  &       \textbf{   89.0 } &          100  &         \textbf{70.5}  \\
    \hline
    SIO w/ FGSM ($\epsilon$=2) &       40.73  &      86.5  &         75.5  &         92.5  &        89.0  &         93.5  &         88.5  &          100  &         70.0  \\
    \hline
    SIO w/ FGSM ($\epsilon$=3) &       40.70  &        86.0  &         75.0  &         91.5  &      \textbf{90.0} &         93.5  &         \textbf{89.0} &          100  &         70.0   \\
    \hline\hline
    \end{tabular}%
    }
  \label{tab:modelscompareAEs}%
\end{table}%

\begin{table}[t]
  \centering
  \caption{The error analysis of AEs generated by our R-FGSM with the SIO for Facebook when attacking different target models.}
  \scalebox{0.9}{
    \begin{tabular}{c|c|c|c|c|c|c}
    \hline\hline
     Models & \multicolumn{2}{c|}{VGG-19bn} & \multicolumn{2}{c|}{ResNet-50} & \multicolumn{2}{c}{Inception-V3} \\
    \hline
    AE sets & OSN Noise &Simulation Error&  OSN Noise  & Simulation     Error& OSN Noise  & Simulation Error\\
    \hline
    
    $\mathbb{A}_s$    & \textbf{22.21 } & \textbf{10.34 }& \textbf{22.46 } & \textbf{11.26 } & \textbf{22.10 } & \textbf{11.19 } \\
    \hline
    $\mathbb{A}_f$    &     30.08  &               19.57   &       22.71  &       11.65  &       29.45  &       15.59       \\
    \hline\hline
    \end{tabular}%
    }
  \label{tab:errorany}%
\end{table}%

\begin{table}[t]
  \centering
  \caption{The average confidence level of Facebook transmitted AEs. These AEs are generated by our R-FGSM with the SIO for Facebook.}
  \scalebox{0.9}{
    \begin{tabular}{c|c|c|c|c|c|c}
    \hline\hline
    \multirow{2}{*}{Image Sets}& \multicolumn{3}{c|}{$\mathrm{ACL}(\mathrm{True},)$} & \multicolumn{3}{c}{$\mathrm{ACL}(\mathrm{False},)$}\\
   \cline{2-7} & VGG-19bn & ResNet-50 & Inception-V3  & VGG-19bn & ResNet-50 & Inception-V3 \\
    \hline
    $\mathbb{O}_s$ & 5.43\% & 6.01\% & 3.39\% & 55.99\% & 52.46\% & 80.24\%\\
    \hline
    $\mathbb{O}_f$ & 65.86\% & 61.08\% & 85.70\% & 18.12\% & 11.46\%  &7.52\%\\
    \hline\hline
    \end{tabular}%
    }
  \label{tab:c5}%
\end{table}%

\begin{table}[t]
  \centering
  \caption{The performance of simulated noise over Facebook, QQ or WeChat by our proposed SIO models and comparative methods. $^+$ Higher is better. $^-$ lower is better.}
    \scalebox{0.85}{
    \begin{tabular}{c|c|c|c|c}
    \hline\hline
    OSNs  & Methods & PSNR$^+$  & SSIM$^+$  & MSE$^-$ \\
    \hline
    & SIO (\textbf{Ours})  &             \textbf{40.63}    &     \textbf{0.9270}   & \textbf{6.85} \\
      Facebook    & ComModel      &              39.42  &         0.8989  &  8.31 \\

        & JPEGR &                39.05  &       0.9051  &    9.94 \\
    \hline
     & SIO (\textbf{Ours}) &           \textbf{37.90 } &       \textbf{0.8723}  & \textbf{11.89} \\
QQ      & ComModel    &              37.03      &     0.8141  &    14.72 \\
       & JPEGR &              34.84        &     0.7839        & 26.38 \\
    \hline
   & SIO (\textbf{Ours})  &              \textbf{33.05 } &       \textbf{0.6536}  & \textbf{206.25} \\
WeChat         & ComModel  &              31.68   &     0.6369     & 223.05 \\
    & JPEGR &                    31.99  &           0.6180        & 219.51 \\
    \hline\hline
    \end{tabular}%
    }
  \label{tab:simu3osn}%
\end{table}%

\subsection{The simulation performance of the SIO}\label{sec:a1}
The simulation performance of SIO is very critical to the ultimate attack effectiveness. In Table \ref{tab:simu3osn}, we give the experimental results on the consistency between SIO and Facebook, WeChat and QQ, and compare them with those of ComModel and JPEGR. Here, the MSE loss, average PSNR and SSIM are calculated with respect to the simulated noise (i.e. $\mathcal{G}_\theta(\mathbf{x})-\mathbf{x}$) and the ground truth OSN noise (i.e. $\mathrm{OSN}(\mathbf{x})-\mathbf{x}$) of 200 AEs, where the associated clean images are randomly selected from the validation set of \texttt{ImageNet}. It can be seen that our SIO models achieve the best simulation results for all three considered OSNs. It also should be noted that the simulation performance of Facebook is generally better than that of QQ and WeChat, for all methods. This is because Facebook operations lead to smaller degradation, which in turn makes the simulation task relatively simple.

\end{document}